\documentclass[11pt]{article}
\usepackage{jheppub}

\pdfoutput=1
\usepackage{amsmath,amssymb}
\usepackage{epsfig}
\usepackage{hyperref}
\usepackage{color}
\usepackage{slashed}
\usepackage{placeins}

\usepackage{amsfonts}
\usepackage{graphicx, rotating}
\usepackage{epstopdf}
\usepackage{latexsym}
\usepackage{rotating}

\usepackage[normalem]{ulem}

\usepackage{subfig}

\usepackage[export]{adjustbox}

\usepackage[shortlabels]{enumitem}

\usepackage[font={small}]{caption}   

\definecolor{myred}{rgb}{0.7, 0, 0}
\definecolor{myblue}{rgb}{0, 0, 0.7}
\definecolor{mygreen}{rgb}{0.04, 0.7, 0.5}

\hypersetup{colorlinks,citecolor=myred,linkcolor=myblue,urlcolor=myblue,linktocpage=true}

 \def\be   {\begin{equation}}   \def\ee   {\end{equation}}
 \def\ba   {\begin{array}}      \def\ea   {\end{array}}
 \def\bea  {\begin{eqnarray}}   \def\eea  {\end{eqnarray}}
 \def\bean {\begin{eqnarray*}}  \def\eean {\end{eqnarray*}}
 
 \def\bry{\begin{array}}
 \def\ery{\end{array}}

\numberwithin{equation}{section}

\begin{document}

\begin{flushright}
\footnotesize
TUM-HEP 1482/23 \\
\end{flushright}
\color{black}

\title{
Hierarchies from Landscape Probability Gradients and Critical Boundaries
}

\author{Oleksii Matsedonskyi}

\affiliation{Technische Universit\"at M\"unchen, Physik-Department, James-Franck-Strasse 1, 85748 Garching, Germany}

\emailAdd{alexey.mtsd@gmail.com}

\abstract{
If the gradient of a probability distribution on a landscape of vacua aligns with the variation of some fundamental parameter, the parameter may be likely to take some non-generic value. Such non-generic values can be associated to critical boundaries, where qualitative changes of the landscape properties happen, or an anthropic bound is located. Assuming the standard volume-weighted and the local probability measures, we discuss ordered landscapes which can produce several types of the aligned probability gradients. The resulting values of the gradients are defined by the ``closeness'' of a given vacuum to the highest- or the lowest-energy vacuum. Using these ingredients we construct a landscape scanning independently the Higgs mass and the cosmological constant (CC). The probability gradient pushes the Higgs mass to its observed value, where a structural change of the landscape takes place, while the CC is chosen anthropically. 
}

\maketitle


\section{Introduction}

Hierarchy problems associated with the Higgs mass and the cosmological constant (CC) remain among the most intriguing challenges of particle physics. 
The most popular approach to the former consists in extending the Standard Model~\cite{Hall:2011aa,Arkani-Hamed:1998jmv,Randall:1999ee,Kaplan:1983fs,Contino:2003ve,Panico:2015jxa} to ensure the technical naturalness of the small Higgs mass.
More recently, an alternative option has emerged~\cite{Dvali:2003br,Dvali:2004tma,Graham:2015cka}, suggesting that the Higgs mass  could be allowed to take much larger values during the cosmological evolution and/or in different patches of the universe, while some type of cosmological selection mechanism would take care of giving a preference to the desired small Higgs mass value. 
This solution shares some degree of conceptual similarity with the anthropic solution to the cosmological constant problem, which assumes the existence of multiple possible vacua with different values of CC.
Furthermore, the cosmological solutions to the Higgs mass and the CC problems are often expected to interfere with each other.

\begin{figure}
\centering
\includegraphics[width=.5\textwidth]{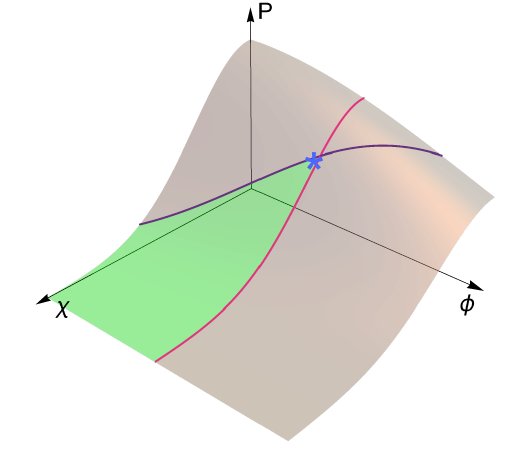}
\caption{Demonstration of the effect of the aligned probability gradients and critical boundaries for a two-dimensional field space case. The probability distribution of $\phi$ and $\chi$ fields values is given by a product of Gaussians with different variances. The critical boundaries $\phi< c_1$ (red line) and $\phi + \chi > c_2$ (purple line) cut out the allowed region (green surface), with the peak of a probability located at the intersection of the critical boundaries (blue star). In the following we will employ a specific example where the $\phi$ field sets the Higgs mass, and the $\phi = c_1$ point corresponds to the SM Higgs mass value; both $\phi$ and $\chi$ also scan the CC, and the critical line approximately given by $\phi + \chi = c_2$ corresponds to the upper anthropic bound on CC.}
\label{fig:sketch}
\end{figure}

In this paper we will follow the second approach to the hierarchy problems and propose a cosmological selection mechanism that can give a statistical preference to apparently unnatural values of parameters, which are scanned by the landscape vacua within a large range.
Concretely, we will demonstrate several cases, where the gradient of probability of different vacua aligns with the variation of the parameter that is scanned. 
These aligned gradients then push the most probable parameter values to their extremes, unless they encounter some type of a critical boundary. 
The latter can be determined either by anthropic arguments, or by structural changes in the landscape properties.

The statistical distribution of the vacua that we analyse depends on the assumed probability measure. We will consider two widely discussed clashing approaches -- the volume-weighted probability measure with a proper-time cutoff~\cite{Linde:1986fd,Linde:1993xx,Linde:1993nz}, and the local measure~\cite{Bousso:2006ev,Susskind:2007pv,Nomura:2012nt}. 
For the volume-weighted measure, we find that the probability gradients are defined by the closeness of a given vacuum to the fastest-inflating vacuum, where the closeness is a function of potential barriers and the vacuum energy variation along the path connecting the two vacua.  
For the local measure, different types of probability gradients can occur depending on the time when the ``sampling" is done, which in our set-up is defined by the duration of the slow-roll inflation.  
Note that we are not aiming here at finding arguments in favour of one measure over the other, and this work is simply intended to analyse possible phenomenological implications of different probability measures.

The proposed approach can be applied to scan several parameters simultaneously. We hence present explicit models scanning independently the Higgs mass and the cosmological constant. The sketch of the corresponding probability distribution over the landscape of vacua is shown in Fig.~\ref{fig:sketch}. The Higgs mass-dependent structural change of the landscape, and the anthropic bound on the CC form critical boundaries on the landscape. The ordered gradient then pushes the values of the scanned parameters to their respective critical values.

This paper is organized as follows. In Section~\ref{sec:globgen} we discuss the volume-weighted probability measure, show how it can lead to the aligned probability gradients in Section~\ref{sec:gradglob}, and then discuss the model selecting the Higgs mass and the cosmological constant in Sections~\ref{sec:mhCCscanglob} and \ref{sec:anthcc}. The main constraints on the model and a simplistic parameter space scan are presented in Section~\ref{sec:mhCCparam}. In  Section~\ref{sec:locgen} we present analogous steps for the case of the local measure. We discuss our results in Section~\ref{sec:conc} and in particular comment on other related approaches of cosmological selection \cite{Geller:2018xvz,Cheung:2018xnu,Giudice:2021viw,Arkani-Hamed:2005zuc,Ghorbani:2019zic}. Appendix~\ref{sec:stoch} contains the derivation of the probability gradients for the volume-weighted measure using the stochastic approach to the field evolution during inflation. Appendix~\ref{sec:bubreh} discusses suppression of non-standard reheating mechanisms in our scenario.

\section{Volume-Weighted Probability Measure}\label{sec:globgen}

Different probability measures give different prescriptions for computing the probability of a given vacuum to be observed (or, in other words, creating an observer in a given type of vacuum).
Volume-weighted probability measures in the simplest case define this probability to be proportional to the overall spacial volume that a given type of vacuum occupies at a fixed proper time~\cite{Linde:1986fd,Linde:1993xx,Linde:1993nz}. Although looking like a quite straightforward way of measuring probabilities, there are important counter-arguments against it. 
First of all, in the regime of eternal inflation that we are focussing on (see the next section for more details), the total volume of vacua is continuously increasing due to the Hubble expansion, which dominates over decays. 
Since this process continues eternally, it is then natural to compare populations of different vacua as a time variable assigned to them goes to infinity. This comparison is ambiguous, as one can use different ways of setting the clocks in different patches of the universe, leading to different probability distributions, see e.g.~\cite{Linde:1993xx}.
Furthermore the simplest version of the volume-weighted probability measure gives rise to a phenomenological problem -- the so-called youngness paradox~\cite{Linde:1994gy,Guth:2000hz,Tegmark:2004qd}. 
To illustrate it let us consider a situation where the eternal inflation is driven by some vacuum with a large CC sourcing the volume expansion with a Hubble rate $H_0$, which can decay with a suppressed rate $\Gamma \ll H_0$ to a zero-CC vacuum, where reheating and creation of observers happen.
The rapid increase of the overall volume of the high-CC vacuum means that at a given time it produces much more new lower-CC universes than it did a finite time ago. The younger habitable universes with an age $t_1$ (measured from their reheating) are preferred compared to those with a higher age $t_2$ by a factor $\sim \exp[3 H_0 (t_2-t_1)]$, where we neglected the milder volume expansion in the low-CC vacuum. Such an exponential preference leads to the conclusion that observing a relatively old universe, such as ours, is highly unlikely~\cite{Linde:1994gy,Guth:2000hz,Tegmark:2004qd}. 

These and other questions related to the volume-weighted measures were a subject of numerous works, see e.g.~\cite{Linde:2007nm,Bousso:2007nd,Linde:2008xf,DeSimone:2008bq,DeSimone:2008if}, and~\cite{Linde:2006nw,Guth:2000hz} for reviews.  
In the following we will assume the version of the volume-weighted measure defined in Ref.~\cite{Linde:2007nm,Linde:2008xf}, the so-called stationary measure with a proper time cutoff, which is argued to be free of the youngness paradox.  
This measure is based on the fact that an eternally inflating universe asymptotically approaches the stationary regime where the volumes of all the vacua have the same scaling $V_i \propto e^{3 H_s t}$. 
The comparison of probabilities of different vacua $i$ (labeled by their features, such as the temperature) happens at equal times $t_{i, \text{measure}}\to \infty$. But the $t_i=0$ point in each type of vacuum is set independently: it is defined as the time when the given vacuum approaches the stationary regime, characterized by $V_i\propto e^{3 H_s t}$. 
Let us apply this prescription to the simple example with two vacua from the beginning of this section.
We then find that 
the lower-CC vacuum is continuously produced from the decays of the high-CC parent, such that the volume of the former  indeed reaches the scaling $V\sim \Gamma \exp[3 H_s t]$, with $H_s \simeq H_0$.
By definition, the younger pocket universes enter the stationary regime earlier than the older ones, their $t_i=0$ is located earlier and so is $t_{i, \text{measure}}$. 
Hence the relative probability of the universes with the age $t_1$ as compared to those with $t_2$ has to include an additional factor $\exp[3 H_s (t_1-t_2)]$ which precisely removes the youngness bias.

In the following we will assume that the contribution of the scanner fields to the vacuum energy is small, so that the stationary regime criterion $V_i\propto e^{3 H_s t}$ is satisfied to a sufficient precision independently of the value of the scanners. 
In such a case there will be no relative difference in $t_i$ for different values of the scanner fields, and therefore, thus defined stationary measure would give the same predictions for the probability distributions over the scanner fields as the standard volume-weighted one. The probability distribution over the inflaton sector fields, which we assume to give the main contribution to $H_s$, instead should be treated with a proper account for the difference of time when the stationary regime is reached.

\subsection{Probability Gradients}\label{sec:gradglob}

\begin{figure}
\centering
\includegraphics[width=.4\textwidth]{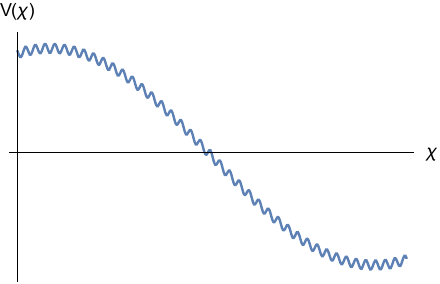}
\hspace{1cm}
\includegraphics[width=.4\textwidth]{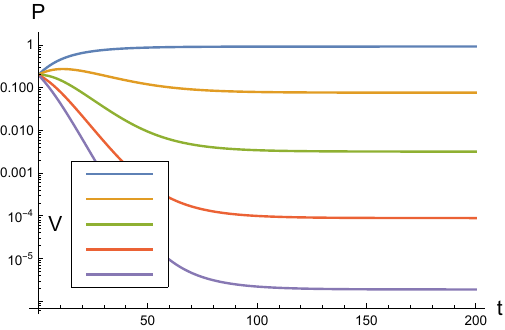}
\caption{{\bf Left:} A sketch of the potential of the one-parameter scanner $\chi$. {\bf Right:} Time evolution of a \emph{normalized} probability distribution for the potential with $N_\chi=5$ local minima, with a vacuum energy ordered as shown in a subplot. After a finite time the distribution enters the stationary regime, with the probability decreasing as one moves away from the highest-$V$ vacuum.}
\label{fig:onefield2cos}
\end{figure}

In this section we will analyse an ordered one-dimensional landscape, scanning a single parameter during inflation, and featuring an aligned probability gradient. 
To this end we introduce a single scanning field $\chi$, whose evolution is  governed by the potential
\bea\label{eq:genscanV}
V(\chi) = 
M_\chi^4 \cos \chi/F_\chi - \mu_\chi^4 \cos \chi/f_\chi + \Lambda_0. 
\eea
We require $F_\chi\gg f_\chi$, so as long as $\mu_\chi^4/f_\chi>M_\chi^4/F_\chi$, this potential features $\sim N_\chi = F_\chi/f_\chi$ local minima, see the sketch in the left panel of Fig.~\ref{fig:onefield2cos}. We also fix $M_\chi \gg \mu_\chi$ for definiteness. To simplify our analysis we assume that the change of the $\chi$ value does not affect the vacuum energy significantly, i.e. 
\be\label{eq:MLhier}
\delta V(\chi) \lesssim M_\chi^4 \ll \Lambda_0,
\ee 
with $\Lambda_0>0$. 
Such type of potentials~(\ref{eq:genscanV}) was introduced in Ref.~\cite{Abbott:1984qf} to scan the CC, and can be generated by the mechanisms of Refs.~\cite{Choi:2015fiu,Kaplan:2015fuy} (see also~\cite{Gupta:2015uea,McAllister:2016vzi}).
The resulting landscape scans the cosmological constant via $V(\chi)$ itself, while the scan of the Higgs mass can be realized by introducing the coupling $\chi h^2$ as will be discussed in detail in the next section.

Important features of the potential~(\ref{eq:genscanV}) are its boundedness and periodicity. The field evolution will never take us outside of the regime with $\delta V(\chi) \ll \Lambda_0$, or where the energy density exceeds the Plank density. Hence we can analyse the $\chi$ evolution without the need to invoke some boundary conditions relying on assumptions about the UV physics.

The time evolution of the physical volumes occupied by each of the $N$ vacua can be analysed using the equation~\cite{Linde:2006nw}
\be\label{eq:Vevoldisc}
\dot P_i = - P_i \sum_{j\ne i} \Gamma_{i\to j} + \sum_{j \ne i} P_j  \Gamma_{j\to i} + 3 H_i P_i,
\ee
where $P_i$ is the volume of the $i$-th local minimum, $H_i^2 =(8\pi/3) V(\chi_i) /m_P^2 $ is its local Hubble parameter, $m_P$ is the Planck mass, $\Gamma_{i\to j}$ is a tunnelling rate of a vacuum $i$ into a vacuum $j$.  
We will assume that the decay rates are small compared to the expansion rates, $\Gamma_i \ll H_i$. 

Let us first assume that the vacua evolve independently, i.e. neglecting small transition rates $\Gamma_{i\to j}$. In this case each vacuum will inflate with its local Hubble rate $H_i$ with the volume growing as $V_i \propto e^{3 H_i t}$. Hence at large times the volume of the highest local minimum (which we call the parent vacuum in the following) exponentially dominates over the volumes of other vacua. Now, turning on small (we will quantify the needed degree of smallness in the following) decay probabilities, we find that evolution of the parent vacuum $i\equiv0$ is governed by the equation 
\be
\dot P_{0} \simeq (3 H_0 - \Gamma_0) P_0,
\ee 
where $\Gamma_0$ is the total parent decay width. This fixes the parent vacuum inflation rate
\be\label{eq:Pp}
P_{0} = C_0 e^{3 H_s t}, \;\; \text{where} \;\; H_s = H_0 - \Gamma_0/3,
\ee
where $C_0$ is a constant. 
The decays of the parent vacuum into the neighbouring vacua, being enhanced by the overall volume of the parent, become sizeable and start affecting their evolution. For simplicity, we only consider the transitions between the nearest neighbours. In this case, the evolution of the second highest minimum $i$ is defined by the equation 
\be\label{eq:lowvacevol}
\dot P_i = 
P_{(i-1)} \, \Gamma_{(i-1)\to i} + 3 H_i P_i,
\ee
where we labeled the patent as $(i-1)$-th vacuum to maintain generality, and neglected the small $i$-th vacuum decay rate.  
Substituting the previously found parent probability $P_0$ from Eq.~(\ref{eq:Pp}) into Eq.~(\ref{eq:lowvacevol}) we find a solution
\be\label{eq:Precur}
P_{i} = C_i  e^{3 H_s t}\;\;\text{where}\;\; C_i = \frac{\Gamma_{(i-1)\to i}}{3(H_s - H_i)} C_{(i-1)}.
\ee
We can then go down the chain of vacua iterating this argument and find that Eq.~(\ref{eq:Precur}) is valid for any pair of neighbours. 
We thus constructed a solution of Eq.~(\ref{eq:Vevoldisc}) where the probability of all the vacua grows exponentially with the rate $H_s$ which is close to the local Hubble expansion rate of the parent vacuum.
This is possible because the missing amount of inflation for the $i$-th vacuum volume to expand with $H_s$, i.e. $H_s - H_i$, is added by the decaying ancestor vacua, as implied by Eq.~(\ref{eq:lowvacevol}). 
Note that in the scenario that we consider the missing amount $H_s - H_i$ is relatively small given the assumption~(\ref{eq:MLhier}). Hence, most of the volume of the $i$-th vacuum is provided by its own volume expansion $3H_i P_i$ rather than by the influx from the $(i-1)$-th vacuum $\Gamma_{(i-1)\to i} P_{(i-1)}$.

The population of each subsequent vacuum is proportional to the factor ${\Gamma_{(i-1)\to i}}/{3(H_s - H_i)}$ which is suppressed by the ancestor's decay width. In this work, for concreteness, we assume that the local curvature of the scanning field potential is subdominant compared to the Hubble rate, i.e. $m_\chi^2 \sim \mu^4/f^2 < H_i^2$. In this regime the tunnelling rate is given by the Hawking-Moss (HM) bounce solution~\cite{Hawking:1981fz,Weinberg:2006pc} with
\be\label{eq:HMbounce}
\Gamma_{j\to i}  \sim H_j  \exp \left[- \frac{8 \pi^2}{3} \frac{\Delta V_B}{H_j^4}\right],
\ee
where $\Delta V_B$ is the barrier height defined as the potential energy difference between the $j$-th minimum and the top of the barrier.

Finally, one should note that if the difference $H_s - H_i$ in the denominator of Eq.~(\ref{eq:Precur}) is smaller than $\Gamma_{(i-1)\to i}$, one formally obtains $P_i > P_{(i-1)}$ and the hierarchical decrease of probability is lost. However, since $H_i$ drops as we move along the chain of minima, it is still possible that the condition $\Gamma_{(i-1)\to i}/(H_s - H_i) \ll 1$ eventually becomes satisfied for some $i$, in which case the subsequent minima can feature a hierarchical suppression.

Assuming uniform Hubble parameter spacing $\Delta H$ between adjacent vacua, and equal downward decay rates $\Gamma_{\downarrow}$ we find by iterating Eq.~(\ref{eq:Precur}) that for the $k$-th vacuum ($k=0$ being the parent)
\be\label{eq:globsinglefieldP}
P_k 
= \left[ \prod_{j=1}^{k} \frac{\Gamma_{\downarrow}}{3 j \Delta H} \right] C_0 e^{3 H_s t}
\simeq \left[ \frac{\Gamma_{\downarrow}}{3 (k/e) (2 \pi k)^{1/2k} \Delta H} \right]^k C_0 e^{3 H_s t}.
\ee
The probability falls monotonically as the $\chi$ value moves away from the global maximum $\chi = 0$. In other words, we have obtained the aligned probability gradient, controlled by the decay rates $\Gamma_\downarrow$ and the Hubble scale variation along the path connecting a given minimum to the parent vacuum. In the next section we will show how such a gradient can be used to select an unnaturally small Higgs mass.

So far we have assumed that the field $\chi$ takes a discrete set of values defined by the local minima of its potential and travels between them due to tunnelling. In fact the $\chi$ field is also a subject to Hubble-induced fluctuations, which are expected to widen the field distribution around the minima and even drive it to the local maxima of the potential. Such an evolution can be described using the stochastic approach of Refs.~\cite{Nambu:1988je,Nambu:1989uf,Sasaki:1988df,Linde:1993xx}. We present the details of this calculation for the potential of the type~(\ref{eq:genscanV})  in Appendix~\ref{sec:stoch} for the case
\be\label{eq:epsdefmain}
\epsilon \equiv \frac 9 {\pi}\frac{H^4}{|m_\chi^2| m_P^2} 
\simeq 
\frac{18}{\pi}\frac{H^4}{\Delta V_B} \frac{f_\chi^2}{m_P^2}
< 1,
\ee
which corresponds to the regime where the effects of the volume growth at the top of the potential barriers are subdominant compared to the classical drift towards the minima which is induced by $\partial_\chi V(\chi)$.  
We find that the resulting relative drop of the probability between adjacent minima is given by Eq.~(\ref{eq:Pminmin}) which is parametrically close to what we have found in the discrete case~(\ref{eq:Precur}), if the limit of small $\epsilon$ is taken. 
Note that $\epsilon$ is automatically small if $\Delta V_B/H^4 \gg 1$, ensuring the hierarchical probability distribution (see Eq.~(\ref{eq:HMbounce})), and $f_\chi \ll m_P$, which we will assume in the following.
Given this conclusion, in the following we will stick to the simpler discrete tunnelling picture. 

One should also mention that in the case $\epsilon >1$ the Hubble expansion favours the field localization around the maxima of the potential~\cite{Giudice:2021viw}. This fact can also be used to create an aligned probability gradient pushing the field towards the global maximum, if the $f$-periodic barriers are absent (at least during inflation). In this case the value of $\epsilon$ is defined by the $F$-periodic part of the potential, with $|m_\chi^2|  \simeq M_\chi^4/F_\chi^2$. 
We will not analyse this option in the following.

\subsection{Higgs Mass and CC Scan}\label{sec:mhCCscanglob}

\begin{figure}
\centering
\includegraphics[width=.4\textwidth]{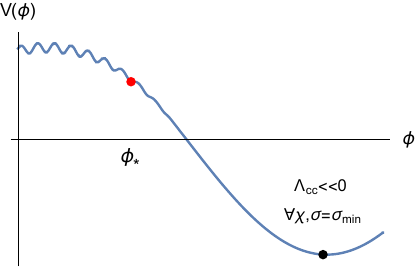}
\caption{A sketch of the potential of the Higgs mass scanner $\phi$. The red dot corresponds to the last local minimum formed by the Higgs-dependent barriers at $h=v_{\text{SM}}$. The absolute minimum (black dot) has a large negative vacuum energy after the end of inflation, i.e. when the inflaton sector fields $\sigma$ settle at their minima.}
\label{fig:phifield2cos}
\end{figure}

Let us now present a model scanning independently two  parameters -- the Higgs mass and the cosmological constant. For now, we assume that  the CC scan is performed by the $\chi$ field with the potential~(\ref{eq:genscanV}) that we have introduced in the previous section, which automatically scans the CC. 
As for the Higgs mass scanner $\phi$ and the Higgs field itself $h$, we assume the following combined potential\footnote{Apart from the terms explicitly written down in Eq.~(2.11), this potential will generically feature additional terms. Most importantly, quantum corrections would necessarily produce Higgs-independent terms with $f_\phi$ periodicity, whose amplitude has to be small enough to not overshadow the sensitivity to the Higgs-dependent term $\mu_\phi^2 h^2 \cos \phi/f_\phi$. This requirement imposes a constrain on $\mu_\phi$ (see Eq.~(2.29)) and also leads to the presence of relatively light new EW-charged states, which can be accessible in near-future collider experiments, which represents an important probe of this scenario~\cite{Graham:2015cka}.}
(see Fig.~\ref{fig:phifield2cos})
\bea\label{eq:Vhphi}
V(\phi, h) &=& 
M_\phi^4 \cos \phi/F_\phi 
- c_1 M_\phi^2 h^2 \cos \phi/F_\phi - c_2 M_\phi^2 h^2 + \frac1 4 \lambda_h h^4 + \mu_\phi^2 h^2 \cos \phi/f_\phi, 
\eea
which is similar, up to a $V\to - V$ reflection, to the relaxion potential proposed in Ref.~\cite{Graham:2015cka}, and further analysed e.g. in Refs.~\cite{Banerjee:2020kww,Chatrchyan:2022dpy,Chatrchyan:2022pcb,Fonseca:2019lmc,Espinosa:2015eda,Choi:2016kke,Banerjee:2018xmn,Fuchs:2020cmm,Gupta:2018wif,Nelson:2017cfv,Fonseca:2017crh}. The main difference with respect to the potential presented in the previous section is the Higgs field dependence of the amplitude of periodic barriers (the last term of Eq.~(\ref{eq:Vhphi})). The Higgs VEV turns on when its $\phi$-dependent squared mass
\be
m_h^2 \simeq - 2 c_2 M_\phi^2 - 2 c_1 M_\phi^2 \cos \phi/F_\phi
\ee
is negative. We assume $c_1 > |c_2|$, so that the Higgs mass is negative around the global maximum of $V(\phi)$, where the $f$-periodic barriers are thus present, while around the global minimum the Higgs VEV and the associated barriers are absent. The last metastable local minimum between the two regimes is located at the critical value $\phi = \phi_*$ defined by the condition $d V(\phi, h)/d \phi = 0$ whose explicit form reads
\be
M_\phi^4/F_\phi  \, \sin \phi_\star/F_\phi = \mu_\phi^2 h^2(\phi_\star) / f_\phi \, \sin \phi_\star/f_\phi.
\ee
The parameters of the potential $V(\phi,h)$ can be fixed such that this last minimum corresponds to the SM value of the Higgs VEV $h(\phi_\star)=v_{\text{SM}}=246$~GeV~\cite{Graham:2015cka}.
After integrating out the Higgs field, the $\phi$ potential takes the approximate form
\be\label{eq:Vphieff}
V(\phi) =  M_\phi^4 \cos \phi/F_\phi + \Theta(\phi_*-\phi) \mu_\phi^2 v_{\text{SM}}^2 \cos \phi/f_\phi,
\ee
where one explicitly sees the structural change happening at the critical value $\phi=\phi_*$.
More precisely, after integrating out the Higgs field we get the term $-m_h^4(\phi)/4\lambda_h \supset \mu_\phi^2 h(\phi)^2 \cos \phi/f_\phi$. 
It can be verified that it leads to the same last minimum condition (2.13) as long as $M_\phi \gg v_{\text{SM}}$. Furthermore, the vacuum energy change produced by $-m_h^4(\phi)/4\lambda_h$ is suppressed compared to that induced by the term $M_\phi^4 \cos \phi/F_\phi$ as long as $(c_1-c_2)$ is mildly suppressed, which will be assumed in the following. For clarity we have also approximated the amplitude of $\cos \phi/f_\phi$ with a constant $\propto v_{\text{SM}}^2$. In the following we will account for the variation of this term with $h$, where relevant.

Other ways to introduce this sort of structural dependence of the scanning field potential on the Higgs mass, such as those studied in Refs.~\cite{Matsedonskyi:2015xta,Matsedonskyi:2017rkq,Arkani-Hamed:2020yna,Khoury:2021zao,TitoDAgnolo:2021pjo}, could also potentially be compatible with the selection mechanisms analysed in this work.

\begin{figure}
\centering
\includegraphics[width=.6\textwidth]{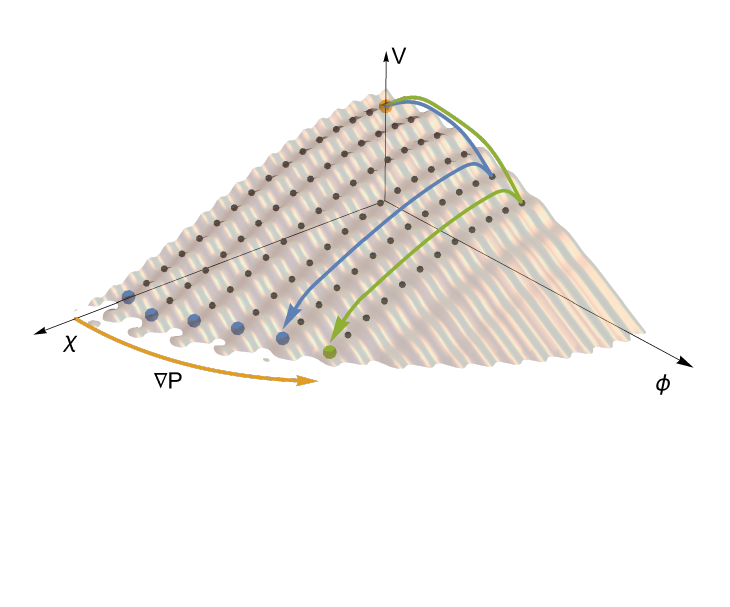}
\vspace{-2.0cm}
\caption{Schematic picture of the potential of the Higgs mass scanner $\phi$ and the CC scanner $\chi$ (grey surface) with a constraint $V\gtrsim0$ after inflation. The green dot shows the vacuum with an anthropic CC and the correct Higgs mass, blue dots show the vacua in the anthropic CC range but with a wrong Higgs mass, an orange dot shows the parent vacuum with a large CC, grey dots are the rest of the local dS minima. Green and blue lines show possible paths from the parent to the vacuum with the correct Higgs mass, and to the vacuum with a wrong Higgs mass respectively. Green line requires one more tunnelling through the $\phi$ barriers and one less tunnelling through the $\chi$ barriers, giving a relative enhancement of $\Gamma_{\phi}/\Gamma_{\chi}$ to the population of the vacua with the correct Higgs mass.
Orange arrow shows the direction of the probability gradient, projected onto the anthropic CC line. }
\label{fig:4}
\end{figure}

We are now set to analyse the combined evolution of the two scanning fields during inflation. The total $\phi,\chi$ potential is schematically depicted in Fig.~\ref{fig:4}. Correspondingly modified probability evolution equation~(\ref{eq:lowvacevol}) reads
\be\label{eq:disc2f}
\dot P_{n_\phi, n_\chi} \simeq  
 \Gamma_{\downarrow \phi} P_{n_\phi-1,n_\chi} + \Gamma_{\downarrow \chi} P_{n_\phi,n_\chi-1}
 +3 H_{n_\phi,n_\chi} P_{n_\phi,n_\chi},
\ee 
where $n_\phi=\phi/(2 \pi f_\phi) $ and $n_\chi = \chi /(2 \pi f_\chi)$ enumerate the $\phi$ and $\chi$ minima and $\Gamma_{\downarrow \phi,\chi}$ give downward tunnelling rates in the $\phi$ and $\chi$ directions. 
Using the assumption~(\ref{eq:MLhier}) we can approximate the Hubble parameter spacings $\Delta H_{\phi}, \Delta H_{\chi}$ between adjacent minima to be uniform and independent, so that $H_{i,j} \simeq H_{0,0} - i \Delta H_{\phi} - j \Delta H_{\chi}$. Then Eq.~(\ref{eq:disc2f}) is solved by
\be\label{eq:pfact}
P_{n_\phi, n_\chi} 
= 
\left[ \prod_{i=1}^{n_\phi} \frac{\Gamma_{\downarrow \phi}}{3 i \Delta H_{\phi}} \right]
\left[ \prod_{j=1}^{n_\chi} \frac{\Gamma_{\downarrow \chi}}{3 j \Delta H_{\chi}} \right] C_0 e^{3 H_s t},
\ee
with the stationary expansion rate being close to the local Hubble rate of the parent vacuum, $H_s \simeq H_{0,0}$.
We thus find that the combined probability distribution is simply given by the product of single-field distributions $P(\phi) \times P(\chi)$.

The landscape defined by $V(\phi,\chi)$ features two critical lines. The first one is located at $\phi=\phi_*$ as discussed above, separating the parts of the landscape with and without the $f_\phi$-periodic local minima.   
The second critical line corresponds to an upper anthropic bound on the cosmological constant, which appears as follows. We assume the presence of an additional sector, collectively denoted $\sigma$, which is responsible for the vacuum energy $\Lambda_0$ driving the eternal inflation. In the course of eternal inflation, the universes with anthropic values of the cosmological constant are continuously born in small fractions of the overall space, where the inflaton sector fields descend from the maximum of their potential, reheating the pocket universes, and eventually reaching the minimum of their potential, where $\Lambda_0$ drops to some value of at most the order of $M_{\phi,\chi}^4$ (see discussion of the inflation sector at the end of this section). As a result, some of the local minima on the $\chi-\phi$ landscape approach the anthropic value of the cosmological constant, some have a CC much above it, and some turn out to be quickly crunching AdS vacua. The part of the $\phi-\chi$ landscape with small- or positive-CC vacua is shown with dots in Fig.~\ref{fig:4}. 

The vacua with a CC closest to the anthropic bound are located on a line in the $\phi-\chi$ plane, which is given by $V(\phi,\chi)^{\text{(today)}} \simeq 0$ and hence $n_\chi$ and $n_\phi$ are related by
\bea\label{eq:V0line}
n_\chi|_{V^\text{(today)}=0}  = \frac{1}{2\pi}\frac{F_\chi}{f_\chi} \arccos \left( const - (M_\phi/M_\chi)^4 \cos (2\pi n_\phi f_\phi/F_\phi ) \right) 
\simeq
- \kappa \,  n_\phi + const,\;\;\;\;\;
\eea
where in the last step for clarity we approximated the anthropic line with a straight line, whose slope $\kappa$ is defined by the slope of the exact line in some typical point $\phi_0,\chi_0$:
\be\label{eq:kappa1}
\kappa = \frac{N_\chi}{N_\phi} \frac{M_\phi^4}{M_\chi^4} \frac{\sin \phi_0/F_\phi}{\sin \chi_0/F_\chi}, \;\;
N_\phi = \frac{F_\phi}{f_\phi}, \;\;
N_\chi = \frac{F_\chi}{f_\chi}.
\ee
The probability distribution~(\ref{eq:pfact}) confined to the line~(\ref{eq:V0line}) reads, in the same approximation, and omitting the common stationary expansion factor as
\be\label{eq:PV0glob}
P(\phi,\chi)|_{V=0} 
\propto  \left( \frac{\Gamma_{\phi}}{3 (n_\phi/e)\Delta H_\phi} \right)^{n_\phi} \left( \frac{\Gamma_{\chi}}{3 (n_\chi/e) \Delta H_\chi} \right)^{n_\chi}
\propto  
\left( 
\frac{e \Gamma_{\phi}}{3 n_\phi \Delta H_\phi}  
	\left( \frac{3 n_\chi \Delta H_\chi}{e \Gamma_{\chi}} \right)^{\kappa} 
\right)^{n_\phi}.
\ee
The resulting distribution~(\ref{eq:PV0glob}) is exponentially peaked at large $\phi$ values as long as
\be\label{eq:mainglob}
\frac{e \Gamma_{\phi}}{3 n_\phi \Delta H_\phi}  
	\left( \frac{3 n_\chi \Delta H_\chi}{e \Gamma_{\chi}} \right)^{\kappa}  \gg 1.
\ee
However there is no local $\phi$ minima at $\phi>\phi_*$, i.e. beyond the first critical boundary. The $\phi=\phi_*$ minimum hence turns out to be the most probable minimum in the anthropic range~\footnote{Note that the vacua with $\phi<\phi_*$ have a lower $n_\phi$ and higher $n_\chi$ which makes the condition~(\ref{eq:mainglob}) easier to satisfy.}. 
 The resulting 2D probability distribution and the two critical lines are schematically shown in Fig.~\ref{fig:sketch}.
We thus demonstrated a mechanism of statistical selection of the small Higgs mass during inflation.

Having introduced the inflationary sector responsible for the scale $\Lambda_0$ we can clarify the question of eternal inflation in our set-up. 
First of all, let us note that even in the absence of the large vacuum energy $\Lambda_0$ the presented $\phi-\chi$ potential is expected to source the eternal inflation with its vacuum energy $\sim M_{\phi,\chi}^4$. Indeed, the suppression of the decay rates compared to expansion in the local minima becomes even stronger when the vacuum energy $\Lambda_0$ is removed, since the decay rates are exponentially sensitive to the Hubble scale. Hence having eternal inflation in such a Higgs mass landscape is not optional. 
The analysed system necessarily enters the stationary regime at large times, with no upper time cutoff. Aiming at finding what is typical for a universe, we are then naturally led to consider the typical, large times, where the stationarity is established.

As for the inflaton sector, we could in principle consider two options. In the first case the inflaton field distribution is maximized at the top of its potential with $V(\sigma) = \Lambda_0$ (which implies that the inflaton sector can support eternal inflation on its own). Therefore for most of the space, the probability distributions of the $\phi$ and $\chi$ fields are formed at some hight $H_s^2 \propto \Lambda_0 \gg M^4_{\phi,\chi}$, as was assumed in the analysis presented above. A small fraction of the overall space, where the inflaton sector fields roll down to the minimum $V(\sigma) \ll \Lambda_0$, contain the reheated universes whose observers can test the values of the Higgs mass. 
We argue in Appendix~\ref{sec:bubreh} that the reverted sequence of events, where the observers are created by the bubble collisions well after the slow-roll inflation, is much less likely.

An alternative option would be to assume that the probability distribution of the inflaton sector fields is maximized around the bottom of their potential. In this case the asymptotic late-time dynamics will be determined by the stationary distribution driven by the potential energy $\sim M_{\phi,\chi}^4$ of the $\phi-\chi$ landscape. This violates our original assumption that the stationary inflation scale is determined by $\Lambda_0$, which was used to derive the field distributions. This also potentially changes the typical ways in which the observers are created, see discussion around Eq.~(\ref{eq:dotPbubble}). 
For these reasons in the following we will assume the first option, with the inflaton sector field distribution maximized in the region with $V \simeq \Lambda_0$.

\subsection{Anthropic Solution for the Cosmological Constant}\label{sec:anthcc}

Let us now discuss the possibility of solving the cosmological constant problem with our model. The $\chi$ field provides a scan of the cosmological constant with the step
\be\label{eq:deltacc}
\Delta\Lambda_{cc \, \chi} \simeq  M^4_{\chi}/N_\chi.
\ee
In order to solve the cosmological constant problem via anthropic selection mechanism this step would have to be at most as large as the today's value of the CC, currently considered to be~\cite{Planck:2015fie}
\be\label{eq:ccobs}
\Lambda_{cc (obs.)} \simeq 10^{-47} \text{GeV}^4,
\ee
which would place a severe constraint on the parameter space.
Furthermore, one should keep in mind that the most probable anthropic value of CC is obtained by a convolution of the field distribution $P(\phi,\chi)$ with the anthropic  factor $A (\Lambda)$, reflecting the probability of emergence of life as a function of CC $\Lambda$. In fact, the field distribution $P(\phi,\chi)$ alone will prefer larger values of CC, since higher CC vacua require less tunnelling steps on the way down from the parent vacuum. Hence $P(\phi,\chi)$ will push the CC towards the upper bound of the anthropic range, which can reach $\sim 10^3 \Lambda_{cc (obs.)}$~\cite{Weinberg:1987dv,Weinberg:1988cp,Bousso:2007kq}.
Therefore, the requirement to maximize $P(\phi,\chi) \times A(\Lambda_{cc})$ close to the observed value~(\ref{eq:ccobs}) implies an additional constraint on the $\chi$ landscape, ensuring a sufficiently flat $P(\chi)$.

In the following we will remove the mentioned constraints, by assuming that $\chi$ only provides a coarse-grained scan of the CC, while a more refined scan allowing to reach the value~(\ref{eq:ccobs}) is performed by a different sector.
The fine-scanning sector can for example be a scaled-down analogue of the $\chi$ sector, or it can even be represented by a non-ordered landscape, in the spirit of Ref.~\cite{Arkani-Hamed:2005zuc}. 
The main features of such a sector should be: a) fine scan of the CC within the interval $\Lambda_{cc \, \text{fine}}$ such that $\Delta\Lambda_{cc \, \chi} \lesssim \Lambda_{cc \, \text{fine}} \ll M_{\chi}^4$, and b) there should be no sizeable probability bias over this additional landscape, which could interfere with the Higgs mass selection, or compete with the anthropic factor $A(\Lambda_{cc})$. 

The only CC-related constraint on the $\chi$ sector that we will impose is the following. Along the critical line $\phi=\phi_*$ the CC is scanned within a range $\sim M_\chi^4$ by the $\chi$ potential. On the other hand, the bare value of the CC is expected to be at least of the order $\text{max}(M_\chi^4,M_\phi^4)$. Since relaxing the CC value by means of the $\chi$ scanner requires the bare value to be at most of the size of the scanning range, we arrive at a conclusion that $M_\phi^4$ should not be much greater than $M_\chi^4$.

\subsection{Parameter Space}\label{sec:mhCCparam}

Let us summarize the main conditions that have to be satisfied by the $\phi, \chi$ landscape in order to produce a statistical preference for the observed small Higgs mass, which we then use to perform a numerical scan of the model parameter space.

\begin{enumerate}

\item
We have worked under assumption of validity of the Hawking-Moss expression of the tunnelling action implying the slow-roll conditions
\bea
m_\phi^2 \simeq \frac{\mu_\phi^2 v^2}{f_\phi^2} < H_s^2, \;\;
m_\chi^2 \simeq \frac{\mu_\chi^4}{f_\chi^2} < H_s^2.
\eea
Note that these assumptions may not be strictly necessary for the efficient selection of the scanned parameters, however we stick to them for simplicity.

\item
We have also assumed that the vacuum energy does not vary significantly across the $\phi - \chi$ landscape, which implies
\be\label{eq:unpertinfl}
M^4_\phi, M^4_\chi \ll \Lambda_0 
\;\Rightarrow\; 
M^4_\phi, M^4_\chi \ll \frac{3}{8\pi} H_s^2 m_P^2.
\ee

\item
We derived our results assuming
\be
\epsilon = \frac{9}{\pi} \frac{H^4}{|m_{\phi,\chi}^2| m_P^2} \ll 1,
\ee
which implies a relative suppression of the Hubble fluctuations-induced probability spread to the areas with higher potential energy.

\item
As was mentioned in the previous section, too small a difference $H_s - H_i$ may prevent the probability suppression in lower-laying minima.  
We thus demand that $10\%$ away from the global maximum the relative probability of each subsequent minimum is suppressed\footnote{Note that, strictly speaking, the statistical preference for the $\phi=\phi_*$ vacuum does not require the lower-laying $\phi$ vacua to feature a hierarchical suppression, see Eq.~(\ref{eq:PV0glob}). A suppression of lower-laying $\chi$ vacua is sufficient. We however demand both distributions to satisfy this condition for simplicity.}. According to Eq.~(\ref{eq:Precur}) this implies
\be\label{eq:parentdom1}
\frac{\Gamma_{\downarrow {\phi,\chi}}}{3 (H_s - H(0.1 \pi F_{\phi,\chi}))} \ll 1,
\ee
where the denominator reads
\be
H_s - H(0.1 \pi F_{\phi,\chi})
= \frac{8\pi}{3}\frac{M_{\phi,\chi}^4 (1-\cos (0.1 \pi))}{2H_s m_P^2}
\simeq 0.2 \frac {M_{\phi,\chi}^4}{ H_s m_P^2}.
\ee

\item
The following condition has to be imposed 
\be\label{eq:lastbarrier}
M_\phi^4/F_\phi = \mu_\phi^2 v_{\text{SM}}^2/f_\phi
\ee
so that the last local $\phi$ minimum approximately corresponds to the SM Higgs VEV value.

Note that the amplitude of the $f_\phi$-periodic term of the $\phi$ potential grows with $h$, saturating at some value which we denote $\mu_{\phi,\text{max}}^4$. The latter can not exceed $\sim (4 \pi)^2 v_{\text{SM}}^4$~\cite{Graham:2015cka}\footnote{This is similar to the QCD axion where the axion potential becomes independent of the Higgs value when the lightest quark mass exceeds the QCD scale~\cite{Kim:2008hd}. For the non-QCD GKR relaxion the role of QCD is played by some other strong sector, while the lightest charged fermion has a mass $\propto h^2$. By dimensional analysis, the saturated magnitude of the $\phi$ potential is $(4 \pi)^2 f_\pi^{\prime 4}$, where the analogue of the QCD pion decay constant $f_\pi^{\prime}$ is required to satisfy $ f_\pi^{\prime} < v_{\text SM}$, see \cite{Graham:2015cka} for details.}. Hence most of the barriers typically have an amplitude which is greater than $\mu_\phi^2 v_{\text{SM}}^2 = \frac{f_\phi}{F_\phi} M_\phi^4$ which would follow from the condition~(\ref{eq:lastbarrier}).
To account for this effect in our numerical scan we will assume a constant $\phi$ barrier amplitude $\mu_{\phi,\text{max}}^4$, and scan over its values within the range $\left[3\frac{f_\phi}{F_\phi} M_\phi^4, (4 \pi)^2 v_{\text{SM}}^4\right]$, where the lower limit is defined by the condition~(\ref{eq:lastbarrier}). The factor of 3 is introduced for computational simplicity, to make sure that the size of the potential barriers, relevant for the computation of the tunneling rate, is dominated by the $f_\phi$-periodic terms, and is simply given by $\Delta V_{B,\phi} \simeq 2 \mu_{\phi,\text{max}}^4$.

Furthermore, the dependence of the barrier amplitude on the Higgs mass would not be overshadowed by quantum corrections as long as~\cite{Espinosa:2015eda}
\be\label{eq:mubound}
\mu_\phi^2  = \frac{f_\phi}{F_\phi} \frac{M_\phi^4}{v_{\text{SM}}^2} < 16 \pi^2 v_{\text{SM}}^2.  
\ee

As for the $\chi$ potential, we simply demand the local minima to be present, and thus impose the condition
\be
M_\chi^4/F_\chi < \frac 1 3 \mu_\chi^4 /f_\chi.
\ee
The factor $\frac 1 3$ is again introduced for computational convenience, so that the height of the barriers separating the $\chi$ minima is given by $\Delta V_{B,\chi} \simeq 2\mu_\chi^4$.

\item
The main Higgs-mass selection criterion~(\ref{eq:mainglob}) reads
\be\label{eq:globmain2}
\frac{e \Gamma_{\phi}}{3 n_\phi \Delta H_\phi}  
	\left( \frac{3 n_\chi \Delta H_\chi}{e \Gamma_{\chi}} \right)^{\kappa}  \gg 1,
\ee
where to derive a parametric estimate of the {\it l.h.s.} we will use $n_\phi = N_\phi/4$, $n_\chi =N_\chi/4$, and
$\kappa = \frac{N_\chi}{N_\phi} \frac{M_\phi^4}{M_\chi^4}$. Note that the expression for $\kappa$ derived in Eq.~(\ref{eq:kappa1}) contains a factor $\sin \phi/F_\phi$ and hence in principle can become very small, removing any suppression of the lower-$\phi$ vacua on the anthropic line. To avoid this we demand that $M_\chi^4 < M_\phi^4/2$. In such a case the anthropic line $M_\phi^4 \cos \phi/F_\phi + M_\chi^4 \cos \chi/F_\chi = const$ does not approach the point with $\sin \phi/F_\phi \simeq 0$ for generic order-one values of $\phi/F_\phi$ and $\chi/F_\chi$ in the point with a correct Higgs mass and anthropic CC.

Furthermore, we demand 
\be
\kappa \geq 1,
\ee
so that a unit decrease of $n_\phi$ along the anthropic line is compensated by at least a unit increase of $n_\chi$, leading to an overall probability drop.

\item
Although the condition (6) ensures a relative statistical enhancement of the desired vacuum over all the others on the anthropic line, the probability of any vacuum on that line is exponentially suppressed compared to that of the parent vacuum. The latter is far outside of the anthropic boundary, and hence it is not suitable for producing the ordinary observers. However it can give rise to the so-called Boltzmann brains (BB) - observers spontaneously created by quantum fluctuations~\cite{Page:2005ur} or by the Hubble-induced fluctuations in the de Sitter background~\cite{Bousso:2006xc}. Ref.~\cite{Bousso:2006xc} estimated the maximal BB creation rate, based on their weight and size only, to be $\Gamma_{\text{BB}} \sim H\exp{10^{-45}}$.  
In the following we will demand that the rate of creating anthropic vacua, which is approximately given by $H_s P(n_\phi,n_\chi)$ with $n_{\phi,\chi} = N_{\phi,\chi}/4$, exceeds the rate of creating BB, which can be estimated as $\Gamma_{\text{BB}} P(0,0)$.  Although that is a very rough estimate of the actual condition, we expect it to be sufficient given the double-exponential suppression of the BB creation rate, which dominates over possible additional factors.

\item
We will now demand that the final anthropic pocket universes produced after inflation are unlikely to feature any domain walls separating regions with different Higgs mass or cosmological constant. 
Let us consider a single-field probability distribution, assuming that the final anthropic vacuum of interest has a probability $P_i$. 
We then separate $P_i$ into a part $P_{i,\text{tun}}$ which has experienced a tunnelling from the $(i-1)$-th minimum within the last 60 e-folds from the reheating time $t_R$, and the part $P_{i,  \text{\sout{tun}}}$ which has not.  
During the last 60 e-folds, the evolution of the part which has not tunnelled is simply described by
\bea
\dot P_{i,\text{\sout{tun}}} &=& 3 H_i \, P_{i,\text{\sout{tun}}}  ,
\eea
where the initial condition is fixed such that  $P_{i,\text{\sout{tun}}}(t_R -t_{60}) = P_{i}(t_R -t_{60})$,
while the part which has undergone a tunnelling is given by $P_{i,  \text{{tun}}} = P_i - P_{i,\text{\sout{tun}}}$. Using the scaling of the stationary regime $P_i = C_i e^{3 H_s t}$ we then find
\be
P_{i,\text{tun}} (t_R) /P_{i,\text{\sout{tun}}} (t_R) = e^{3 \cdot 60 \frac{H_s-H_i}{H_i}} - 1.
\ee
Hence most of the volume of the $i$-th vacuum  will be concentrated in domains of space which 
are unlikely to feature domain walls within our today's observational reach as long as
\be\label{eq:nodomwall}
\frac{H_s-H_i}{H_i} \ll 1/180. 
\ee
This condition is similar to the condition (2), since ${(H_s-H_i)}/{H_i} \simeq {(H_s^2-H_i^2)}/{2H_i^2} \simeq {M_{\phi,\chi}^4}/{2\Lambda_0}$.

\end{enumerate}

\begin{figure}
\centering
\includegraphics[width=.31\textwidth]{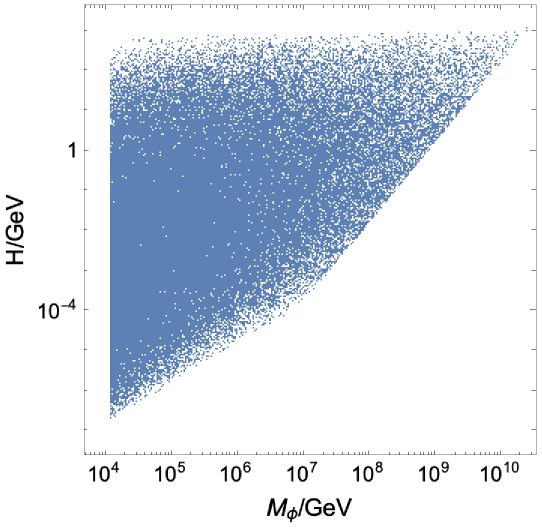}
\hspace{0.1cm}
\includegraphics[width=.31\textwidth]{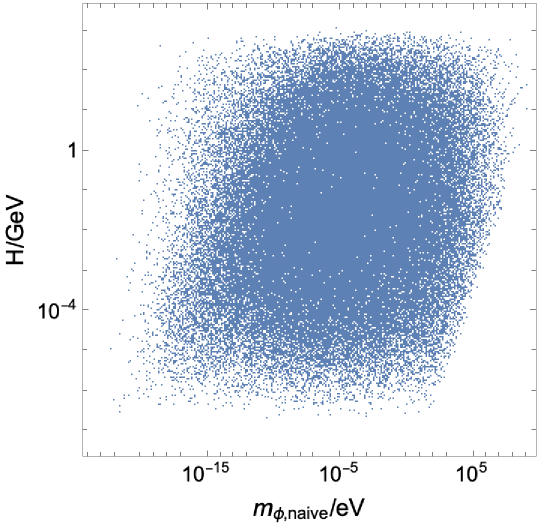}
\hspace{0.1cm}
\includegraphics[width=.312\textwidth]{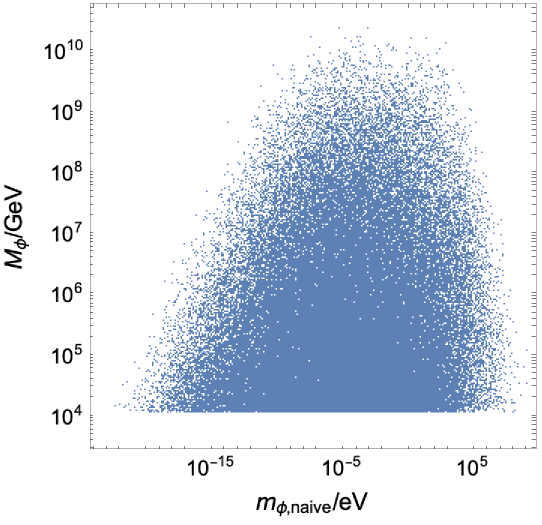}
\caption{Scatter plots of the allowed parameter space points for the case of volume-weighted measure, in terms of $M_\phi$, Hubble scale of inflation $H$, and $m_{\phi, \text{naive}} = \mu_\phi v_{\text{SM}}/f_\phi$. The latter is a naive estimate of the $\phi$ scanner mass today, which can however exceed its actual value~\cite{Banerjee:2020kww}.}
\label{fig:scanglob}
\end{figure}

We present the results of a numerical scan of the model parameter space in Fig.~\ref{fig:scanglob}. 
We perform the scan in the range $m_P > f_{\phi}(f_{\chi})> M_{\phi} (M_{\chi}) > 10$~TeV, $\mu_{\chi} < 0.1 M_{\chi}$, and $M_\chi^4>10^{-2} M_\phi^4$ (as needed for the CC scan, see Section~\ref{sec:anthcc}), while the parameters $\mu_{\phi,\text{max}}, H_s, F_\phi, F_\chi$ are scanned within the maximal range allowed by the listed bounds. 
We find the cutoff of the Higgs mass scanner field to lie within $M_\phi \lesssim 10^{10}$~GeV, which corresponds to the scale where new physics coupled to the standard model is located (see end of Section~\ref{sec:conc} for more details). The inflationary Hubble scale is constrained to be $H\lesssim 100$~GeV. These two bounds can be understood as follows.
The inflationary Hubble scale has to stay below ${\Delta V_B}^{1/4} \sim \mu_{\phi,\text{max}}$ in order to produce a hierarchical suppression between the $\phi$ vacua, as required by the condition~(\ref{eq:parentdom1}), where $\mu_{\phi,\text{max}} \lesssim v_{\text{SM}}$. This results, roughly, in $H \lesssim 100$~GeV. Furthermore, requiring the scanning sector to represent a subdominant fraction of the overall inflationary potential, which is encoded in the conditions~(\ref{eq:unpertinfl}), (\ref{eq:nodomwall}), we arrive at the bound $M_\phi \lesssim \sqrt{m_P H}$, leading to   $M_\phi \lesssim 10^{10}$~GeV when accounting for the upper bound on the Hubble scale. 

Most of the parameter space has $H\ll v$, where inflation-induced fluctuations of the Higgs field can be neglected and the Higgs field is expected to stay around its classical minimum, as was assumed in this analysis. Note however, that large Higgs displacements away from the minimum would not necessarily be a problem for the considered scenario. For example, large non-vanishing Higgs VEV during inflation would imply the $\phi$ barriers to be present all the time during inflation, still potentially leading to the discussed pattern of probability gradients. The critical line around $h = v_{\text{SM}}$ would only appear after inflation, when the Higgs field settles in its minimum. This might still lead all the field space on one side of the critical line into AdS, before observers could form. As for the case of vanishing Higgs VEV during inflation, see discussion at the end of Section~\ref{sec:gradglob}.

\section{Local Probability Measure}\label{sec:locgen}

The proponents of the local probability measures~\cite{Bousso:2006ev,Susskind:2007pv,Nomura:2012nt} (also sometimes called holographic or causal) argue that the global structure of the universe, containing multiple causally disconnected regions, is not suitable for computing probabilities. Instead, one should consider the cosmological history accessible to a hypothetical observer following a worldline passing from the formation of the universe, following transitions between different vacua, until it finally ends in a crunching anti-de Sitter vacuum (a sink). 
The probability distribution of observable vacua is then determined by the probability of this observer to enter a given vacuum on the way from the initial vacuum to the sink. This distribution of vacua is then weighted by some type of anthropic factor.

Restricting to a single worldline observer makes the volume expansion beyond the observer's horizon irrelevant. 
Furthermore, in a landscape of vacua containing sinks, most of worldline observer histories will end up in a sink within a finite amount of time. Hence the local approach discards both fundamental aspects related to the volume-weighted measure, which tries to capture the possibly eternal (in the future) existence of the universe and its continuous volume growth. 
Thus the youngness paradox, stemming from a large volume of eternally produced younger vacua, is avoided.

A more in-depth introduction to the local measures and the argumentation in their support can be found e.g. in Refs.~\cite{Bousso:2006ev,Susskind:2007pv,Nomura:2012nt}. As was stated in the introduction, we will not try to compare the plausibility of different measures, and hence proceed directly to the phenomenological implications relevant for this paper, namely the aligned probability gradients.

\subsection{Probability Gradients}

We again start with the case of a single scanning field $\chi$, with the potential defined by Eq.~(\ref{eq:genscanV}) and depicted in Fig.~\ref{fig:onefield2cos}, and we will make the same assumptions about its parameters. In particular, we impose the condition $M_\chi^4 \ll \Lambda_0$ which allows to separate the evolution of the inflationary sector from that of the scanning sector. This condition implies that before the end of $\Lambda_0$-induced inflation all the $\chi$ vacua have a positive vacuum energy. We again assume that at the end of the slow-roll inflation the vacuum energy drops by $\sim \Lambda_0$, so that a fraction of the $\chi$ landscape turns into AdS minima.

The time evolution of a worldline can be described with an equation similar to the one used in the volume-weighted case~(\ref{eq:Vevoldisc}), after removing the term reflecting the volume growth~\cite{Linde:2006nw}
\be\label{eq:Vevoldiscloc}
\dot P_i = - P_i \sum_{j\ne i} \Gamma_{i\to j} + \sum_{j \ne i} P_j  \Gamma_{j\to i}, 
\ee
where $P_i$ is the probability of a worldline to be in the $i$-th vacuum at a given time, and $\Gamma_{j\to i}$ are the transition rates which we assume to correspond to the HM tunnelling.
Note that also in this case the Hubble-size  fluctuations can take the field outside of  local minima. This process can again be described using the stochastic approach, leading to a transition rate which is parametrically close to the one predicted by the discrete tunnelling via the HM bounce~\cite{Chatrchyan:2022pcb}.

Unlike the case of the volume-weighted measures, most of the transitions between vacua in this case happen within a finite time, until most of them fall in AdS sinks (which appear after the end of the slow-roll inflation). 
Hence instead of considering the distribution in the $t\to \infty$ limit we need to study Eq.~(\ref{eq:Vevoldiscloc}) at finite times. There are several typical timescales separating qualitatively different regimes of the probability evolution. For simplicity we again only consider the transitions between the nearest neighbours, with equal upward tunnelling rates $\Gamma_{\uparrow}$ and equal downward ones $ \Gamma_{\downarrow}$.  
Correspondingly, the following distinct time scales can be singled out:

\begin{enumerate}[A)]

\item
$t \ll \Gamma_{\uparrow,\downarrow}^{-1}$:

In this regime the dynamics responsible for transitions between the vacua would not have time to substantially alter the initial probability distribution of the $\chi$ field $P(t=0)$. 
Aligned probability gradients, even if exist, are solely determined by the initial conditions. As an example we will consider the initial condition proposed in Ref.~\cite{Linde:1983mx,Vilenkin:1984wp} for a quantum creation of the universe
\be\label{eq:qcreation}
P(t=0)
\propto \exp\left[-\frac {3}{8} \frac{m_P^4}{V(\chi)}\right] 
\propto \exp\left[\frac {8 \pi^2}{3} \frac{V(\chi)}{H^4}\right],
\ee
where in the second step we assumed a small relative vacuum energy variation across the $\chi$ landscape. The distribution~(\ref{eq:qcreation}) assigns the highest probability to the highest-energy vacuum in the $\chi$ landscape, and creates an ordered probability gradient at early times before the dynamics has any effect. However this distribution is controlled by the vacuum energy only, and therefore, in the analysed two-dimensional landscape it would not give any preference to the needed Higgs mass value once the value of the CC is fixed by the anthropic considerations.

\item
$t \lesssim \Gamma_{\downarrow}^{-1}$:

In this case the evolution defined by the controllable vacuum dynamics can substantially alter the initial distribution. The general solution of Eq.~(\ref{eq:Vevoldiscloc}) reads
\be\label{eq:chaindecay}
P = \exp[\kappa t] P(t=0) \;, \text{with} \; \kappa_{i j} =  \Gamma_{j \to i} - \delta_{i j} \sum_k \Gamma_{j \to k}.
\ee
To be more specific we stick with the previously introduced initial condition~(\ref{eq:qcreation}). It favours the highest-energy vacua, which would then decay, thereby populating the lower-laying states. If this process dominates the probability, the occupancy of the $i$-th vacuum (with $i=0$ corresponding to the parent) is given by an expansion of Eq.~(\ref{eq:chaindecay})
\be\label{eq:locallowtgrad}
P_i \simeq \frac{1}{i!} (\kappa t)^i P(t=0) \simeq \frac{1}{i!} (\Gamma_{\downarrow} t)^i,
\ee
where we used the fact that at least $i$ downward decays are needed in this case, and took $P(t=0,i=0)\simeq1$.
Eq.~(\ref{eq:locallowtgrad}) thus defines the aligned probability gradient, determined by an interplay of initial conditions and controllable dynamics, i.e. decays with the rate $\Gamma_{\downarrow}$.

On the other hand, at large enough times with
\be\label{eq:regime2up}
\Gamma_{\downarrow} t \simeq 1
\ee 
the probability of the highest vacuum significantly depletes, and the monotonically decreasing distribution gets distorted. Eq.~(\ref{eq:regime2up}) thus sets the upper time cutoff of this regime.

\begin{figure}
\centering
\includegraphics[width=.5\textwidth]{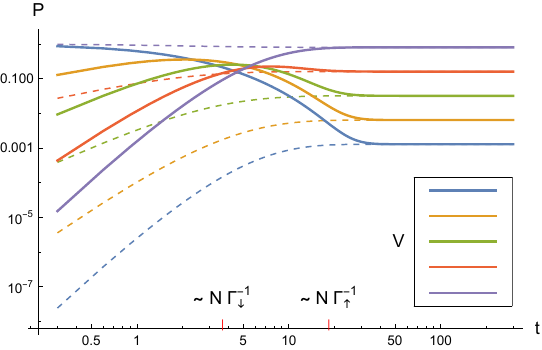}
\caption{Probability evolution for $N=5$ ordered vacua, for the initial probability distribution localized in the highest vacuum (solid lines) and the lowest one (dashed lines), in the case of the local measure. $\Gamma_{\downarrow}$, $\Gamma_{\uparrow}$ are respectively downward and upward decay rates. The vacuum energy decreases from blue to purple, as shown in a sub-plot. The distribution passes from the initial-conditions-dominated regime to the equilibrium, as discussed in the main text.}
\label{fig:localchi1field}
\end{figure}

\item
$t > N_\chi \, \Gamma_{\uparrow}^{-1}$: 

With the suppressed upward tunnelling most of $P$ would be eventually concentrated at the lowest energy level, however the upward transitions will create a spread of probability around the ground state. In this regime one can expect to reach a dynamical equilibrium where downward and upward transitions balance each other, so that 
\be\label{eq:locequil}
0 = \dot P_i = - P_i \sum_{j\ne i} \Gamma_{i\to j} + \sum_{j \ne i} P_j  \Gamma_{j\to i}.
\ee
This type of static distribution for the local measure was found in Ref.~\cite{Bousso:2006ev} for the so-called cyclic landscapes, i.e. the ones with no AdS sinks\footnote{Note that $P$ in the notations of Ref.~\cite{Bousso:2006ev} is the integrated incoming probability current rather than the total probability.}.
In the presence of sinks, instead, the probability of anthropically-relevant dS vacua would gradually deplete due to decays into crunching sinks, without reaching the exact equilibrium. We are eventually interested in the landscape with AdS vacua, however if the vacuum energy controlled by the inflation sector stays large for a sufficiently long time, the system of $\chi$ vacua can be effectively considered as cyclic for that period since all of them have positive vacuum energy.  

Eq.~(\ref{eq:locequil}) is solved, up to an overall normalization, by
\be\label{eq:loceqsol}
P_i 
\propto \exp\left[\frac {3}{8} \frac{m_P^4}{V(\chi_i)}\right]
\propto  \exp\left[-\frac {8 \pi^2}{3} \frac{V(\chi_i)}{H^4}\right].
\ee
This can be verified by using the explicit form of the HM tunnelling rate, or using the Lee-Weinberg argument~\cite{Lee:1987qc} ${\Gamma_{\downarrow}}/{\Gamma_{\uparrow}} 
\simeq  \exp\left[\frac {8 \pi^2}{3} \frac{|\Delta V_1|}{H^4}\right]$, where $\Delta V_1$ is the potential energy difference between two minima.

Eq.~(\ref{eq:loceqsol}) predicts hierarchically distributed probabilities for sufficiently separated vacuum energies. Note however that the distribution is solely sensitive to the vacuum energy, while the barriers' properties or the field distance do not play any role.  This is similar to the case of the initial condition~(\ref{eq:qcreation}) assumed in the regime A, although the dependence on the vacuum energy is opposite.

\end{enumerate}

The reheating at the end of the slow-roll inflation produces observers.  Hence the reheating time $t = t_R$ defines the time when the probability distribution $P$ should be computed, selecting one of the listed above regimes. 
In Fig.~\ref{fig:localchi1field} we show the time evolution for the system with $N_\chi=5$ vacua, obtained by solving Eq.~(\ref{eq:Vevoldiscloc}) numerically. We use two different initial conditions, preferring the highest vacuum, with the initial distribution $P(t=0)  = \delta_{i 0}$ (solid lines), or the lowest one with $P(t=0)  = \delta_{i 4}$ (dashed lines).

\subsection{Higgs Mass and CC Scan}\label{sec:locmhcc}

We now introduce the Higgs mass scanner $\phi$ in the same manner as was done in Section~\ref{sec:mhCCscanglob}. The overall potential of the two fields is shown in Fig.~\ref{fig:4}. We will analyse the combined Higgs mass and CC scan in the regime B assuming the initial probability distribution~(\ref{eq:qcreation}). In the regimes A and C instead, the distribution is determined by the vacuum energy alone, and therefore an independent scan of the Higgs mass can not be performed in a straightforward manner. 
It is then also natural to assume the initial condition~(\ref{eq:qcreation}) for the inflaton sector fields, which means that they are initially localized at the top of their potential. The typical worldline trajectories will then start with a period of $\Lambda_0$-dominated inflation, following through a period of slow roll and the usual reheating as the inflaton sector evolves down to the minimum of its potential energy. 

As before, we will assume that the considered landscape only provides a rough scan of the CC, up to the value specified in Eq.~(\ref{eq:deltacc}), while the final scan is performed by an additional sector of the type discussed in Section~\ref{sec:anthcc}. We expect that the evolution in the fine-scan sector can be decoupled from that of the $\phi$ and $\chi$ fields, as long as the former satisfies rather simple criteria similar to those listed in Section~\ref{sec:anthcc}.

In the analysed regime, the lower-laying $\phi$ and $\chi$ vacua are accessed from the dominant top vacuum. Assuming that the Hubble scale variation can be neglected in the expression for the decay rates, the probability distribution for the two fields factorizes (see Eq.~(\ref{eq:locallowtgrad})) 
\be\label{eq:Pphichiloc}
P_{n_\phi, n_\chi} 
= 
\frac{1}{n_\phi !} \left[  \Gamma_{\downarrow \phi} t_R \right]^{n_\phi}
\frac{1}{n_\chi !} \left[  \Gamma_{\downarrow \chi} t_R \right]^{n_\chi},
\ee
where $t_R$ is the time of reheating.
The projection of $P_{n_\phi, n_\chi} $ on the anthropic CC line~(\ref{eq:V0line}) has the form
\be\label{eq:PV0loc}
P(\phi,\chi)|_{V=0} 
\simeq  
\left( \frac{\Gamma_{\phi} t_R}{n_\phi/e} \right)^{n_\phi} 
\left( \frac{\Gamma_{\chi} t_R}{n_\chi/e} \right)^{n_\chi}
\propto  
\left( 
\frac{\Gamma_{\phi} t_R}{n_\phi/e}  
	\left( \frac{n_\chi/e}{\Gamma_{\chi} t_R} \right)^{\kappa} 
\right)^{n_\phi},
\ee
and the exponential preference for larger $\phi$ values is ensured by the condition
\be\label{eq:mainloc}
\frac{\Gamma_{\phi} t_R}{n_\phi/e}  
	\left( \frac{n_\chi/e}{\Gamma_{\chi} t_R}  \right)^{\kappa}  \gg 1,
\ee
which will push $\phi$ to the critical boundary corresponding to the observed Higgs mass, similarly to the case of the volume-weighted measure.

Let us now summarize the main constraints on the model. 
First of all, we impose the selection condition~(\ref{eq:mainloc}) with $n_{\phi,\chi} = N_{\phi,\chi}/4$, $\kappa \geq 1$ (as defined under Eq.~(\ref{eq:globmain2})) and $M_\chi^4<M_\phi^4/2$ for the same reason as before. 
The main constraints on the model include (1), (2), (5) of the volume-weighted measure case discussed in Section~\ref{sec:mhCCparam}. Also similarly to the volume-weighted measure, we argue in Appendix~\ref{sec:bubreh} that the reheating from the bubble walls collisions is irrelevant.
Furthermore, there are additional constraints specific to the local measure:

\begin{enumerate}

\item
As was argued in the previous section, the regime B is realized for
\be
t_R \lesssim  \Gamma^{-1}_{\downarrow  \phi},  \Gamma^{-1}_{\downarrow  \chi}.
\ee
We will impose this bound on both $\chi$ and $\phi$ fields for definiteness, however the needed preference for larger $\phi$ values is only expected to be stronger if the $\phi$ distribution exits the regime B and moves to C.

\item
We would like to avoid formation of domain walls in the visible part of the universe.
Hence the contribution to the population of the anthropic minima from the tunnellings happening within the last $\sim 60$ e-folds of inflation has to be subdominant. For a single-field probability of the $k$-th minimum this condition implies
\be\label{eq:domwall}
\frac{P_k({\text{tunnel within }}t_R - t_{60} < t < t_R)}{P_k({\text{tunnel within }} t< t_R - t_{60})}
\simeq \frac{k t_{60}}{t_R}
\ll 1,
\ee
where $t_{60}=60 H^{-1}$, and $k\sim N_{\phi,\chi}/4$ for a minimum on the anthropic line.

\item
Since the initial condition $P(t=0)$ is degenerate with the vacuum energy, it gives the same probability for any Higgs mass value on the anthropic line. Thus we demand that the initial distribution~(\ref{eq:qcreation}) is completely erased by the vacuum transition dynamics~(\ref{eq:chaindecay}), $P(n_{\phi,\chi}) > P (t=0,n_{\phi,\chi})$, which implies
\bea\label{eq:P0overdyn}
\frac{1}{n_\phi!}\left[{\Gamma_{\phi\downarrow} t_R} \right]^{n_\phi} &>& \exp\left[-\frac {8 \pi^2}{3} \frac{V(0) - V(n_{\phi})}
{H^4}\right], \\
\frac{1}{n_\chi!}\left[{\Gamma_{\chi\downarrow} t_R} \right]^{n_\chi} &>& \exp\left[-\frac {8 \pi^2}{3} \frac{V(0) - V(n_{\chi})}
{H^4}\right].
\eea
In the following we will again set $n_{\phi,\chi} \simeq N_{\phi,\chi}/4$ to obtain a parametric estimate of the bound, and take $V(0) - V(n_{\phi,\chi}) \simeq M_{\phi,\chi}^4$.

\item
As for the Boltzmann brain condition, we will now demand that the integrated BB probability $\sim \Gamma_{\text{BB}} t_R$ be less than the probability of a typical vacuum on the anthropic line, $\sim P(N_\phi/4,N_\chi/4)$ at the time of reheating.

\end{enumerate}

\begin{figure}
\centering
\includegraphics[width=.313\textwidth]{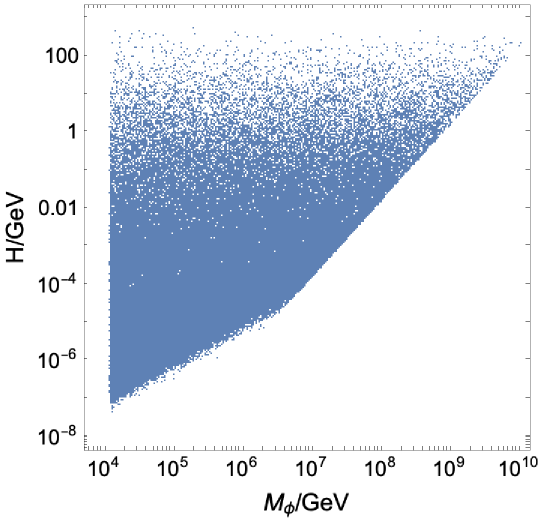}
\hspace{0.1cm}
\includegraphics[width=.314\textwidth]{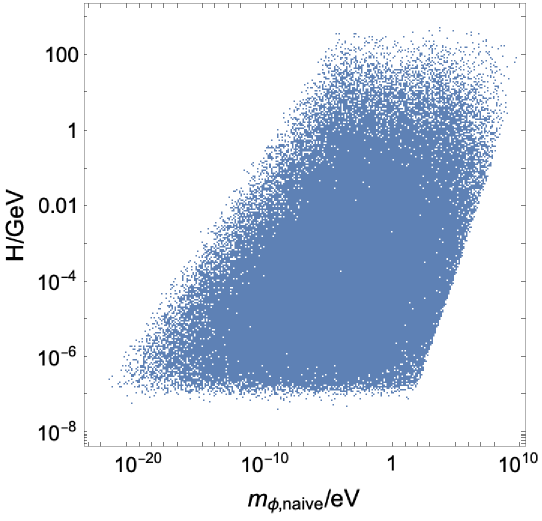}
\hspace{0.1cm}
\includegraphics[width=.327\textwidth]{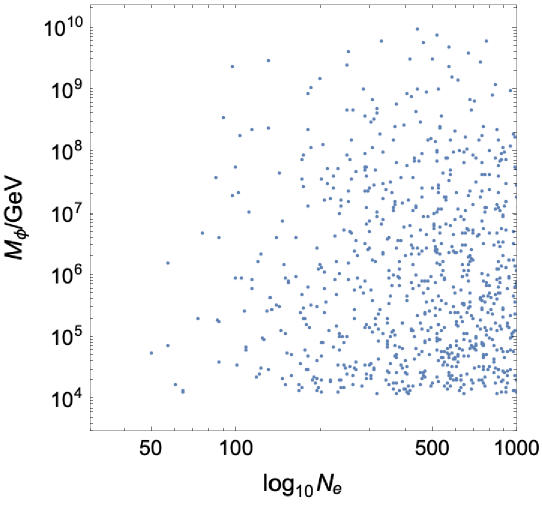}
\caption{Scatter plots of the allowed parameter space points for the case of the local measure, in terms of $M_\phi$, $H$, $m_{\phi, \text{naive}} = \mu_\phi v_{\text{SM}}/f_\phi$ and the number of e-foldings of inflation $N_e = H t_R$.
In the third plot we do not show points with $\log_{10} N_e >10^3$ to allow for a sufficient visibility of the lower-$N_e$ end.}
\label{fig:scanloc}
\end{figure}

We present the results of the parameter space scan in Fig.~\ref{fig:scanloc}, for the same scan range as in the volume-weighted case. In addition, we scan over the duration of inflation within the entire range allowed by the mentioned bounds.
The resulting bounds on the SM sector cutoff $M_\phi \lesssim 10^{10}$~GeV and the Hubble rate $H\lesssim 10^2$~GeV are similar to the volume-weighted case. The upper bound on $H$ is dictated by the combination of the  constraints (1) and (2), which force $\Gamma_{\phi} \ll 1$ and hence $H\lesssim v_{\text{SM}}$.  The condition $M^4_\phi \ll \frac{3}{8\pi} H^2 m_P^2$ then leads to the mentioned upper bound on $M_\phi$. 

The bound  (3)  leads to a severe constraint on the duration of inflation, as can be seen in the third plot in Fig.~\ref{fig:scanloc}~\footnote{See e.g. Ref.~\cite{Kitajima:2019ibn} for inflationary models providing large $N_e$.}. 
Note also that the actual time of inflation needed to establish the probability distribution that we used may be longer than $t_R$ which we obtained in our scan. The reason is that while discussing the field distributions we were assuming instantaneous transitions between different minima. In practice, the HM bounce describes the tunnelling to the top of the potential barrier, from where the field has to travel classically or by stochastic fluctuations to the next local minimum, which would entail some time delay.

\section{Discussion}\label{sec:conc}

We have analysed mechanisms creating probability distribution gradients which are aligned with a variation of some field in a landscape. 
If such a field is coupled to some operator, the latter is scanned, with a statistics driven by the aligned gradient. The probability then can naturally peak close to the extreme values of the scanned parameter. These extreme values can in particular be defined by anthropic arguments, or by the structural changes in the landscape properties.
We applied this mechanism to scan the cosmological constant and the Higgs mass, assuming the anthropic bound on the former, and implementing a structural change of the landscape around the SM value of the Higgs mass, thus selecting the latter statistically.  

Discussion of a landscape probability distribution generically requires assuming some specific probability measure. However, various proposed measures are known to be able to produce rather different phenomenological predictions. In this regard, our approach here was to demonstrate the fact that two orthogonal, but widely advocated prescriptions for measuring the probability can in principle feature the needed aligned gradients, although with different properties. 
On the other hand, in the absence of a consensus about theoretical motivations of different measures, the phenomenological implications (such as the values of the Higgs mass and the CC, together with other possible experimental or observational signatures) could serve to find out which of the measures is physically relevant.

We have discussed the measure rewarding volume expansion of different vacua, and the local measure which is agnostic to expansion. In the regime of eternal inflation, relevant for our model, the former measure typically predicts the landscape to enter a stationary regime with a probability distribution which is independent of the initial conditions. Given that the eternal inflation does not feature any physical time cutoff, it is natural to study the probability distributions at large times when the stationarity is established. As a result, the gradients, and the preferred values of the fundamental parameters can be determined with only minimal assumptions about the initial conditions for our universe.
The local measure, insensitive to possible eternal self-reproduction of the universe, instead naturally features finite time scales which can be relevant in determining the probability distribution. In our set-up this time scale is set by the reheating time of the slow-roll inflation sector. It defines the moment when the ordinary observers can start being produced in each of the landscape vacua.  
The observed probability distribution depends substantially on this time, thus connecting the properties of the slow-roll inflation sector with the observed values of the fundamental parameters. 
We analysed in detail the intermediate time regime, where the statistics is determined by an interplay of initial conditions and the controllable dynamics of the landscape.

There is a number of other proposals for selecting the Higgs mass and/or the CC in a dynamical/statistical manner~\cite{Geller:2018xvz,Cheung:2018xnu,Giudice:2021viw,Khoury:2021grg,Arkani-Hamed:2005zuc,Ghorbani:2019zic,Strumia:2020bdy,TitoDAgnolo:2021pjo,TitoDAgnolo:2021nhd,Csaki:2020zqz,Bloch:2019bvc,Matsedonskyi:2017rkq,Kaloper:2023kua,Hook:2016mqo}. In relation to the topic of this paper, we would like to mention in particular the following approaches. 
Concentrating on the volume-weighted measure, Refs.~\cite{Geller:2018xvz,Cheung:2018xnu} analysed landscapes with the critical value of the scanned parameter corresponding to the maximum of the potential energy, hence providing it with the maximal volume enhancement. 
In the case of the volume-weighted measure, the mechanism discussed in the present work does not assume the critical value to correspond to the potential energy maximum.
This fact in particular allows to straightforwardly select a small Higgs mass and a small CC simultaneously. Furthermore, the volume enhancement is not used at all in the case of the local measure that we present. On the other hand, unlike Refs.~\cite{Geller:2018xvz,Cheung:2018xnu}, the anthropic factor plays a major role in the concrete model that we analysed, coming as a price of solving (or relaxing) the CC hierarchy problem.
Volume enhancement during the (typically eternal) inflation for a single scanner case was also studied in Ref.~\cite{Giudice:2021viw}, where, differently from our scenario, the critical point was also given a maximal probability over the landscape, as a consequence of being located at the junction of two phases of the landscape.

We have performed a simplistic and non-exhaustive analysis of the parameter space of the model which scans the Higgs mass and the CC with both considered measures and found that the maximal Higgs mass cutoff is $\sim 10^{10}$~GeV. This cutoff corresponds to the scale where the new electroweak-charged physics is supposed to reside, which can thus be well beyond the near future collider reach.
However, implementations of landscapes scanning the Higgs mass naturally lead to the Higgs interactions with the scanner fields, which can be probed experimentally. In particular, the scanner studied in this paper has properties similar to the relaxion, with promising signatures in particle physics experiments~\cite{Flacke:2016szy,Frugiuele:2018coc,Banerjee:2019epw,Budnik:2019olh,Banerjee:2020kww,Fuchs:2020cmm}. Similarly to the GKR relaxion the $\phi$ field of the presented scenario is expected to undergo a dynamical tuning of its mass leading to improved detection prospects~\cite{Banerjee:2020kww,Fuchs:2020cmm}.
Furthermore, 
the critical dependence of the $\phi$ landscape barriers on the Higgs field implies new TeV-scale particles accessible in foreseeable future experiments~\cite{Graham:2015cka}.
One of the generic challenges of this paradigm (which can not be counted as a factor making the discussed mechanisms less plausible) is related to the reconstruction of the global landscape structure from the particle physics experiments, dealing with the expansion around a single local minimum. However one may hope to get more insights from the observational imprints of the early universe dynamics~\cite{Banerjee:2021oeu}, and astrophysics~\cite{Banerjee:2019epw,Hook:2019pbh,Balkin:2021zfd,Balkin:2021wea,Servant:2023mwt,Budnik:2020nwz,Balkin:2023xtr,Budker:2023sex}, where parts of the global vacuum structure can in principle be probed.  Another interesting approach is related with a search for the indirect signs of near-criticality of the Higgs potential~\cite{Steingasser:2023ugv}.

\vspace{1cm}
{\bf Acknowledgements}

We thank Alexander Westphal, Gilad Perez, Hyungjin Kim, Andi Weiler, Ben Allanach and Jim Talbert for discussions. 
This work has been partially supported by the Collaborative Research Center SFB1258, the Munich Institute for Astro- and Particle Physics (MIAPP), the Excellence Cluster ORIGINS, which is funded by the Deutsche Forschungsgemeinschaft (DFG, German Research Foundation) under Germany’s Excellence Strategy – EXC-2094-390783311, and STFC HEP Theory Consolidated grant ST/T000694/1.

\appendix

\section{Stochastic Approach}\label{sec:stoch}

When quantizing a light scalar field in dS background one finds that after some time $t\gtrsim H^{-1}$ the field oscillations get stretched to the Hubble size and freeze, effectively creating a stochastic classical background field. In this case the field evolution can be described by a stochastic Langevin equation. One can alternatively pass to the description in terms of a field distribution, experiencing the drift induced by the scalar potential, diffusion induced by the dS background, and growth due to the Hubble expansion. In the slow-roll approximation $\ddot \phi \ll 3 H \dot \phi$ the evolution of the volume-weighted distribution is controlled by the following modified Fokker-Planck equation
\be\label{eq:evolcontV}
\dot P = \frac{\partial}{\partial \phi} \left( \frac{H^{3(1-\beta)}}{8 \pi^2} \frac{\partial}{\partial \phi} (H^{3\beta} P)\right) + \frac{\partial}{\partial \phi} \left( \frac{V'}{3H}P\right) + 3 H P,
\ee
where e.g. Ref.~\cite{Linde:1993xx} used $\beta = 1/2$, although other order-one values are also used in literature. The exact value however will not matter for the level of approximation we will work at.
Eq.~(\ref{eq:evolcontV}) provides a complementary description of the probability evolution, compared to the one provided by the discrete evolution equation~(\ref{eq:Vevoldisc}). The latter includes the quantum tunnelling processes, while the former includes classical slow rolling and Hubble jumps, although the two descriptions are argued to be related~\cite{Linde:2006nw}. We refer the reader to Ref.~\cite{Giudice:2021viw} for a recent in-depth analysis of various properties of solutions of Eq.~(\ref{eq:evolcontV}).

We will approximate the complete solution to the ``wiggly'' potentials of the type~(\ref{eq:genscanV}) and~(\ref{eq:Vphieff}) with a collection of solutions for the quadratic potentials stitched together.

\subsection{Quadratic Potential}

Let us first consider a single patch with a quadratic potential
\be\label{eq:Vquad}
V = \Lambda + \frac1 2 m^2 \phi^2.
\ee
We will search for the solution in the form 
\be
P (\phi,t) = \sum_{H_i} e^{3 H_i t} P_{H_i}(\phi).
\ee
The allowed values of $H_i$ have to be fixed by the initial or boundary conditions which we discuss in the following. We will eventually be interested in the solution with the maximal allowed $H_i$ which we call $H_s$ -- the stationary inflation rate. This solution has the highest expansion rate and therefore dominates at large times. 
The potential~(\ref{eq:Vquad}) substituted into the evolution equation~(\ref{eq:evolcontV}) gives
\bea \label{eq:evolcontVquad}
P''_{H_s} 
+ \left\{ \frac 3 8 \frac{m_P^4 m^2}{\Lambda^2}  \phi \right\} P'_{H_s} 
+ \left\{ -\frac 3 8 \frac{m_P^2}{\Lambda^2}  \left(24 \pi \Lambda^{1/2} (\Lambda_{s}^{1/2} - \Lambda^{1/2} ) - m_P^2 m^2 \right) + \frac {9 \pi}{4} \frac{ m_P^2  m^2}{\Lambda^2} \phi^2\right\}  P_{H_s} =0\, \nonumber \\
\eea
where we only retained the leading terms in $\frac{V(\phi)-V(0)}{V(0)}$ and $\Lambda/m_P^4$ expansion, and denoted $\frac{8 \pi}{3}\Lambda_{s} = H_s^2 m_P^2$. 
The solutions of Eq.~(\ref{eq:evolcontVquad}) can be expressed using the parabolic cylinder functions $D$
\be\label{eq:gensol}
P_\nu = 
\exp\left[-A \phi^2\right] 
\big\{ 
{\bf c_{+}} D_{\nu}\left[ B \phi \right] + {\bf c_-} D_{\nu}\left[- B \phi \right]
\big\},
\ee
where we traded $H_s$ for $\nu(H_s)$, which is defined together with other parameters as 
\bea
A \phi^2 
&=&
\frac{4 \pi^2}{3}\frac{V(\phi)-V(0)}{H(0)^4},\label{eq:Adef}\\
B \phi 
&=& 
\left\{4 \frac{4 \pi^2}{3}\frac{|V(\phi)-V(0)|}{H(0)^4} \sqrt{1-\frac 9 \pi \frac{H(0)^4}{m^2 m_P^2}} \right\}^{1/2} \text{sign}[\phi],\\
\nu
&=&
\frac{9(H(0)^2 - H_s^2)+ m^2}{2|m^2| \sqrt{1-\frac 9 \pi \frac{H(0)^4}{m^2 m_P^2}}} - \frac 1 2 .
\label{eq:defnu}
\eea
Note that $|A| \phi^2$ and $B^2 \phi^2$ have the same parametric form as the HM bounce action for the $\phi$ tunnelling to the top of the barrier, and hence we will assume them to be large. 

The two chosen basis functions of Eq.~(\ref{eq:gensol}) become degenerate for non-negative integer $\nu$, where a different choice of basis is needed, but this will never be relevant for our purposes.

We will only consider the case 
\be
\epsilon \equiv \frac 9 {\pi}\frac{H^4}{|m^2| m_P^2} \ll 1,
\ee
so that the arguments and the order of the $D$ functions remain real. For our specific potentials such as~(\ref{eq:genscanV}), this parameter can be rewritten as $\epsilon \simeq \frac{18 H^4}{\pi \Delta V_B} \frac{f^2}{m_P^2}$ where $\Delta V_B$ is the barrier height. The first factor is typically small, as needed to suppress the tunnelling between the different minima, and therefore $\epsilon$ is automatically small for $f \ll m_P$.

At large absolute values of the arguments $B \phi$, 
the $D$ functions have the asymptotic form
\bea\label{eq:Dasymp}
 D_{\nu}(x) 
&\underset{\begin{subarray}{c}
x \to \infty
\end{subarray}}{\longrightarrow}&
|x|^\nu e^{-x^2/4} 
\\
&\underset{\begin{subarray}{c}
x \to -\infty
\end{subarray}}{\longrightarrow}&
(-1)^{\nu} |x|^\nu e^{-x^2/4} 
+  \frac{\sqrt{2\pi}}{\Gamma[-\nu]} |x|^{-\nu-1} \,  e^{x^2/4}\nonumber \\
&\underset{\begin{subarray}{c}
x \to 0
\end{subarray}}{\longrightarrow}&
\frac{2^{\nu/2} \sqrt{\pi}}{\Gamma[\frac{1-\nu}{2}]} \nonumber
\eea

\subsection{Probability Current}

Integrating the equation~(\ref{eq:evolcontV}) over an interval $(\phi_1,\phi_2)$ we obtain
\be\label{eq:FPint}
3 H_s \int_{\phi_1}^{\phi_2} P_\nu(\phi)  \, d \phi = \int_{\phi_1}^{\phi_2} 3 H(\phi) P_\nu(\phi)  \, d \phi  + J(\phi_1) - J(\phi_2),
\ee
where $J$ is a probability current in the $\phi$ direction 
\be
J = - \frac{H^3}{8 \pi^2} \left[ H^{-3 \beta } \partial_\phi (H^{3\beta}P_\nu(\phi)) + \frac{8 \pi^2}{3}\frac {V'}{H^4} P_\nu(\phi) \right],
\ee
with its two components being Hubble-induced diffusion and classical rolling respectively. The equation~(\ref{eq:FPint}) has the same form as the discrete Eq.~(\ref{eq:Vevoldisc}), with discrete quantities substituted with the integrals over $\phi$ patches, and the probability change due to tunnelling substituted with the currents.
We also find that the stationary expansion rate is given by 
\be\label{eq:Hsint}
H_s = \frac{\int_{\phi_1}^{\phi_2} H(\phi) P_\nu(\phi) d \phi}{\int_{\phi_1}^{\phi_2}  P_\nu(\phi) d \phi} + \frac{J(\phi_1) - J(\phi_2)}{3\int_{\phi_1}^{\phi_2} P_\nu(\phi) d \phi}.
\ee
In particular, for the isolated field range (such that there is no flow at the endpoints, $J(\phi_{1,2})=0$) one sees that  $H_s$ is just the $P$-averaged value of the local Hubble parameter. On the other hand, the actual expansion rate $H_s$ which is greater (lower) than the average local expansion rate requires positive (negative) net flow of probability into the given field patch.  The explicit form of the probability current for the quadratic potential in $\epsilon \ll 1$ expansion is
\bea
 J 
=
-\left(\frac{H^3}{8\pi^2} \right) e^{-A \phi^2}
&& \{
{\bf c_+}  
\left( 
2A (1-\text{sign}[m^2] + \epsilon/2)\phi D_{\nu}[B \phi]  + \nu B D_{\nu-1}[B \phi] 
\right) \\
&&+ 
{\bf c_-} 
\left( 
2A (1-\text{sign}[m^2] + \epsilon/2)\phi D_{\nu}[-B \phi]  - \nu B D_{\nu-1}[-B \phi] 
\right)  
\}, \nonumber\\
\eea 
where we used the relation $D_\nu[x]'=-\frac x 2 D_\nu[x] + \nu D_{\nu-1}[x]$ and neglected the term $\propto \beta \Lambda/m_P^4$.

\subsection{Boundary Conditions} \label{sec:BC}

{\bf $\partial_t$ matching:}

When combining two solutions we need to equate their time derivatives, i.e. $H_s$ on either side. Using Eq.~(\ref{eq:defnu}), and assuming equal in absolute value masses, one then finds a relation between the values of $\nu$ in a maximum and an adjacent minimum, $\hat \nu$ and $\check \nu$ respectively (in the following we use ``~$\hat{}$~" and ``~$\check{}$~" for the quantities related to maxima and minima, where needed)
\be\label{eq:dtmatch}
 \check \nu \simeq (\hat \nu + 1) - \frac 9 2 \frac{\hat H(0)^2- \check H(0)^2}{|m^2|},
\ee
where we neglected $\epsilon$.
Note that for our specific potential the last term in Eq.~(\ref{eq:dtmatch}) can be rewritten as
\be\label{eq:deltaHmaxmin}
\frac 9 2 \frac{\hat H(0)^2- \check H(0)^2}{|m^2|}  \simeq 24 \pi \frac{f^2}{m_P^2}.
\ee
When working with sub-Planckian $f$, the quantity~(\ref{eq:deltaHmaxmin}) will typically be small, 
 so that $\check \nu \simeq \hat \nu + 1$. As follows from~Eq.~(\ref{eq:Hsint}), the highest possible expansion rate $H_s$ for the isolated field range can not exceed the maximal local value of the expansion rate $H(\phi)$. Hence the global maximum of the potential has $\hat \nu \gtrsim -1$, with $\hat \nu \simeq -1$, if allowed, giving the maximal possible stationary expansion rate (see Eq.~(\ref{eq:defnu})). It then follows from Eq.~(\ref{eq:dtmatch}), that the highest local minimum, adjacent to the global maximum, will have $\check \nu \sim 0$ for maximal $H_s$.

\vspace{0.3cm}
{\bf $P$ matching:} 

The continuity of probability gives one matching condition per boundary.  In the following it will be useful to have the expressions for the probability for the previously found extreme values of $\hat \nu, \check \nu$. Around minima and maxima we find, respectively
\bea\label{eq:gensolasym}
\check P_{\nu}(\phi) 
&\underset{\begin{subarray}{c}
\phi\to \infty
\end{subarray}}{\longrightarrow}&
({\bf c}_+ + (-1)^\nu {\bf c}_-)|B\phi|^\nu  e^{-2A \phi^2 (1-\epsilon/4)} 
+ {\bf c}_- \frac{\sqrt{2\pi}}{\Gamma[-\nu]} |B \phi|^{-\nu-1} \,  e^{- \epsilon A \phi^2/2}
\\
&\underset{\begin{subarray}{c}
\nu \to 0 \\
\phi\to \infty
\end{subarray}}{\longrightarrow}&
({\bf c}_+ + {\bf c}_-)  e^{-2A \phi^2 (1-\epsilon/4)} 
- {\bf c}_- \sqrt{2\pi} \nu |B \phi|^{-1} \,  e^{- \epsilon A \phi^2/2}
\\
&\underset{\begin{subarray}{c}
\phi \to 0
\end{subarray}}{\longrightarrow}&
({\bf c}_+ + {\bf c}_-) \frac{2^{\nu/2}\sqrt \pi}{\Gamma[\frac{1-\nu}{2}]}
\;\;\underset{\begin{subarray}{c}
\nu \to 0 \\
\end{subarray}}{\longrightarrow}\;\;
({\bf c}_+ + {\bf c}_-) \\
\hat P_\nu(\phi)
&\underset{\begin{subarray}{c}
\phi\to \infty
\end{subarray}}{\longrightarrow}&
({\bf c}_+ + (-1)^\nu {\bf c}_-)|B\phi|^\nu  e^{- \epsilon |A| \phi^2/2} 
+ {\bf c}_- \frac{\sqrt{2\pi}}{\Gamma[-\nu]} |B \phi|^{-\nu-1} \,  e^{2|A| \phi^2 (1+\epsilon/4)}
\\
&\underset{\begin{subarray}{c}
\nu \to -1 \\
\phi\to \infty
\end{subarray}}{\longrightarrow}&
({\bf c}_+ - {\bf c}_-)  |B \phi|^{-1} e^{- \epsilon |A| \phi^2/2} +  {\bf c}_- \sqrt{2\pi}   e^{2|A| \phi^2 (1+\epsilon/4)} \label{eq:Pmaxapprox} \\
&\underset{\begin{subarray}{c}
\phi \to 0
\end{subarray}}{\longrightarrow}&
({\bf c}_+ + {\bf c}_-) \frac{2^{\nu/2}\sqrt \pi}{\Gamma[\frac{1-\nu}{2}]}
\;\;\underset{\begin{subarray}{c}
\nu \to -1
\end{subarray}}{\longrightarrow} \;\;
\sqrt{\pi/2}({\bf c}_+ + {\bf c}_-) 
\eea
The expressions in the limit $\phi \to -\infty$ are obtained by exchanging ${\bf c_+} \leftrightarrow {\bf c_-}$.

\vspace{0.3cm}
{\bf $J$ matching:}

We also need to match $\partial_\phi$ derivatives at the boundaries. We will do that by equating the currents $J$, which is equivalent to equating $P(\phi)'$ in the case of a potential differentiable across the boundary (which is the case for us). However equality of $J$ also holds when the derivative of a potential is discontinuous~\cite{Giudice:2021viw}.
Another useful property of $J$ is that its sign shows the direction of the probability flow. The explicit form of the currents in the minima and maxima is, respectively
\bea\label{eq:jminmax}
\left(\frac {8\pi^2} {H^3} \right) \check J_\nu
%
%
&\simeq&
-   A \epsilon \phi  \check P(\phi)
-   e^{-A \phi^2}
  \nu B \left({\bf c_+} D_{\nu-1}[B \phi] - {\bf c_-} D_{\nu-1}[-B \phi]\right) \\
%
%
&\underset{\begin{subarray}{c}
\phi\to \infty
\end{subarray}}{\longrightarrow}&
-   A \epsilon |\phi| \check P(\phi)
-  
({\bf c_+}+ (-1)^\nu {\bf c_-}) \nu B |\phi B|^{\nu-1} e^{-2A \phi^2 (1-\epsilon/4)} 
+    {\bf c_-} \frac{\sqrt{2\pi} \nu B}{\Gamma[1-\nu]}  |\phi B|^{-\nu} e^{-\epsilon A \phi^2 /2} 
 \nonumber\\
%
%
&\underset{\begin{subarray}{c}
\nu \to 0 \\
\phi\to \infty
\end{subarray}}{\longrightarrow}&
-   A \epsilon |\phi| \check P(\phi)
- ({\bf c_+}+ {\bf c_-}) \nu B |\phi B|^{-1} e^{-2A \phi^2 (1-\epsilon/4)} 
+   {\bf c_-} \sqrt{2\pi} \nu B e^{-\epsilon A \phi^2 /2}\\
%
%
&\underset{\begin{subarray}{c}
\phi\to 0
\end{subarray}}{\longrightarrow}&
 \frac{2^{\frac{\nu-1}{2}}\sqrt \pi}{\Gamma[\frac{2-\nu}{2}]}\, \nu B ({\bf c_-} - {\bf c_+}) 
\;\;\underset{\begin{subarray}{c}
\nu \to 0
\end{subarray}}{\longrightarrow} \;\;
\sqrt{\pi/2} \, \nu B ({\bf c_-} - {\bf c_+}) 
\label{eq:jmin0}
\\
 \left(\frac{8\pi^2}{H^3} \right) \hat J_\nu
%
%
&\simeq&
-|A| \epsilon \phi \hat P(\phi)+
  e^{|A| \phi^2}
B 
\{
{\bf c}_+ D_{\nu+1}[B \phi] - {\bf c}_- D_{\nu+1}[ - B \phi]
\} 
\\
%
%
&\underset{\begin{subarray}{c}
\phi\to \infty
\end{subarray}}{\longrightarrow}&
-|A| \epsilon |\phi| \hat P(\phi)
+  ({\bf c}_+ +(-1)^\nu {\bf c}_- ) B |\phi B|^{\nu+1} e^{- \epsilon |A| \phi^2/2}
- {\bf c}_- \frac{\sqrt{2 \pi}B}{\Gamma[-\nu-1]} |B\phi|^{-\nu-2} e^{2|A| \phi^2 (1+\epsilon/4)}\nonumber\\
%
%
&\underset{\begin{subarray}{c}
\nu \to -1 \\
\phi\to \infty
\end{subarray}}{\longrightarrow}&
-|A| \epsilon |\phi| \hat P(\phi)
+  ({\bf c}_+ - {\bf c}_- ) B e^{- \epsilon |A| \phi^2/2}
+ {\bf c}_- {\sqrt{2 \pi}B(\nu+1)} |B\phi|^{-1} e^{2|A| \phi^2 (1+\epsilon/4)}\\
%
%
&\underset{\begin{subarray}{c}
\phi\to 0
\end{subarray}}{\longrightarrow}&
\frac{2^{\frac{\nu+1}{2}}\sqrt \pi}{\Gamma[\frac{-\nu}{2}]}\, B ({\bf c}_+ - {\bf c}_- ) 
\;\;\underset{\begin{subarray}{c}
\nu \to -1 
\end{subarray}}{\longrightarrow}\;\;
B ({\bf c}_+ - {\bf c}_- ) 
\label{eq:jmax0}
\eea 
where we applied the relation $D_{\nu+1}[x] = x D_{\nu}[x] - \nu D_{\nu-1}[x]$.
The asymptotes in the limit $\phi \to -\infty$ are obtained by exchanging ${\bf c_+} \leftrightarrow {\bf c_-}$ and reverting the overall signs of $\check J, \hat J$.

\subsection{Chain of Quadratic Potentials}

\begin{figure}
\centering
\includegraphics[width=.4\textwidth]{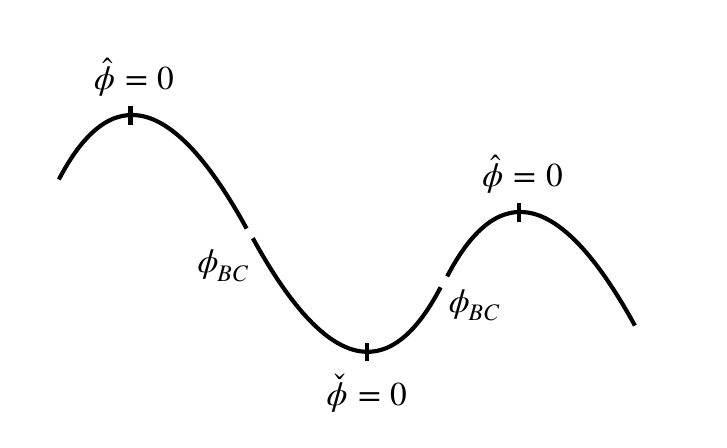}
\caption{Matching scheme of quadratic potentials.}
\label{fig:sketch_appendix}
\end{figure}

Let us consider a periodic landscape consisting of a series of alternating minima and maxima defined by quadratic potentials with masses equal in absolute value, and with a matching happening at the equal distance from the nearest maximum and minimum. At the matching points where boundary conditions (BC) are imposed we assume
\be
\left[(B\phi)^2 \right]_{\text{BC}} 
, \left[ 4 |A| \phi^2 \right]_{\text{BC}}  \gg 1.
\ee 
We further assume for definiteness that the landscape is periodic and reflection-symmetric with respect to its global minimum and a global maximum.
Because of this symmetry, we can concentrate on the field range between the global maximum and the global minimum. 
For each individual quadratic patch we set $\phi=0$ at the center. Hence e.g. for a maximum and a minimum to the right of it, the common boundary is located at $\phi > 0$ for the maximum, and at $\phi <0$ for the minimum.
In order to introduce an overall tilt to the potential (to mimic the effect of the $2\pi F$-periodic $\cos$) we assume the matching between each minimum and the next maximum happening at a slightly lower $|\phi|$ than the matching between each maximum and the following minimum. This matching scheme is shown in Fig.~\ref{fig:sketch_appendix}.

Sufficiently far from the global maximum one expects the local Hubble rate to drop significantly below $H_s$, so that in all minima $\check \nu \ll 0$, and $\hat \nu \ll -1$, which we will assume for now. 
One can notice that for same-order coefficients ${\bf c}_+ \sim {\bf c}_-$ and order-one values of $\hat \nu, \check \nu$, the values of $P$ and $J$ are dominated by the larger exponents. As we show below, this results in the values of $J/P$ being independent of ${\bf c}$-coefficients. Let us consider a maximum and a minimum next to it.  Using the asymptotic expansions at $\phi \to \pm \infty$ we then find that for generic $\nu$
\bea
\left(\frac {8\pi^2} {H^3} \right) \hat J/ \hat P 
&\underset{\begin{subarray}{c}
\phi\to \infty \\
{\bf \hat c}_+ \sim {\bf \hat c}_-
\end{subarray}}{\longrightarrow}&
\frac {1}{|\phi|} \left\{
-\frac 9 2 \frac{\hat H(0)^2 - \hat H(\phi)^2}{|m^2|}    
+ (\hat \nu + 1)
\right\}, \label{eq:jpmax}
\\
\left(\frac {8\pi^2} {H^3} \right) \check J/ \check P
&\underset{\begin{subarray}{c}
\phi\to -\infty\\
{\bf \check c}_+ \sim {\bf \check c}_-
\end{subarray}}{\longrightarrow}&
\frac {1}{|\phi|} \left\{
\frac 9 2 \frac{\check H(\phi)^2 - \check H(0)^2}{|m^2|}  
+   |\check B \phi|^2 
\right\},\label{eq:jpmin}
\eea
where we used the relation
\be
|A|\epsilon \phi^2 \simeq \frac 9 2 \frac{|H(\phi)^2 - H(0)^2|}{|m^2|} .
\ee 
The independence of Eqs.~(\ref{eq:jpmax}), (\ref{eq:jpmin}) on ${\bf c}$ coefficients for ${\bf c}_+ \sim {\bf c}_-$ and generic $\check \nu < 0$, $\hat \nu<-1$ means that the boundary values of $J/P$ can not be tuned to match. 
The matching can only be done if one of the ${\bf  c}$ coefficients controlling the larger exponents is highly suppressed, hence the $J/P$ value becomes sensitive to the relative size of the ${\bf  c}$ coefficients and hence is adjustable. One way to make a matching is by tuning down ${\bf \hat c}_-$: 
\be
\left| \frac{{\bf \hat c}_-}{{\bf \hat c}_+}  \right|
\lesssim   
-\frac{\Gamma[-\hat \nu]}{\sqrt{2\pi}} |\hat B \phi_{\text{BC}}|^{2 \hat \nu + 1} e^{-2|A| \phi^2_{\text{BC}} (1+\epsilon/2)},
\ee
such that $\hat J/ \hat P (\phi \to \infty)$ is brought to the size of $\check J/ \check P (\phi \to - \infty)$. 
Alternatively, one could make the BC tunable by suppressing ${\bf \check c}_+$:
\be\label{eq:mintun}
\left| \frac{{\bf \check c}_+}{{\bf \check c}_-}  \right|
\simeq  
\frac{\Gamma[1-\check \nu]}{\sqrt{2\pi} } |\check B \phi_{\text{BC}}|^{2 \check \nu -1} e^{-2 A \phi^2_{\text{BC}} (1-\epsilon/2)}.
\ee

Matching a minimum with a subsequent maximum, i.e. $\check J/ \check P(\phi \to \infty)$ with $\hat J/ \hat P(\phi \to -\infty)$, would require analogously to tune down ${\bf \hat c}_+$ or ${\bf \check c}_-$.
For a given quadratic patch one can choose either ${\bf c}_+ \ll {\bf c}_-$ or ${\bf c}_- \ll {\bf c}_+$, to satisfy the BC on its left or right side respectively, while the BC on the opposite side should be satisfied by tuning in the adjacent quadratic patch. 
Hence, 
all the tunings have to be made on one side, by tuning all of the ${\bf \check c}_-,{\bf \hat c}_-$ or all of the ${\bf \check c}_+,{\bf \hat c}_+$. On the other hand, the absolute minimum, by symmetry, can only have ${\bf \check c}_- = {\bf \check c}_+$. Hence ${\bf \hat c}_-$ of the maximum to the left of the absolute minimum has to be tuned, and also all the other ${\bf c}_-$. This is true until $\check \nu$ approaches zero, which we will discuss later on. 

This one-sided tuning of ${\bf c}_-$ coefficients implies that in the local origins of the minima and maxima the currents are positive, i.e. go in the direction of the absolute minimum, see Eqs.~(\ref{eq:jmin0}),~(\ref{eq:jmax0}).
Furthermore, the relative suppression of the coefficients implies a suppression of the probabilities in each successive quadratic patch. By equating the probabilities at the boundaries we find that the probability at the top of the barrier $\hat P_i(0)$ is suppressed with respect to the probability at the minimum to the left of it $\hat P_i(0)$ by 
\be\label{eq:Pminmax}
\frac{\hat P_i(0)}{\check P_i(0)} 
\,\simeq\,
\frac{2^{ \hat \nu_i /2} \Gamma[\frac{1-\check \nu_i}{2}] }{ 2^{ \check \nu_i /2} \Gamma[\frac{1-\hat \nu_i}{2}]}
\frac{{\bf \hat c}_{i,+}}{{\bf \check c}_{i,+}}
\,\simeq\,
\frac{2^{ \hat \nu_i /2} \Gamma[\frac{1-\check \nu_i}{2}] }{ 2^{ \check \nu_i /2} \Gamma[\frac{1-\hat \nu_i}{2}]}
\frac{\Gamma[-\hat \nu_i]}{\sqrt{2\pi}} |B \phi_{\text{BC}}|^{\check \nu_i + \hat \nu_i +1}
e^{-\frac{8 \pi^2}{3}\frac{\Delta V_B}{H^4}},
\ee
where we approximated $B = \hat B = \check B$ in the sub-exponential factor and $H\simeq \hat H_i(0) \simeq \check H_i(0)$ in the exponent. $\Delta V_B=\hat V(0) - \check V(0)$ is the height of the potential barrier. The terms linear in $\epsilon$ cancel from the exponent, leaving an expression similar to the HM tunnelling rate. 
Note however that there is an $\epsilon^2$ correction, which we neglected in the exponent.

By matching with the next minimum, we further find that the probability density in the minimum to the right of the barrier $\check P_{i+1}(0)$ is suppressed compared to the probability in the left-side minimum $\check P_{i}(0)$, by
\be\label{eq:Pminmin}
\frac{\check P_{i+1}(0)}{\check P_i(0)} 
\,\simeq\,
\frac{{\bf \check c}_{i+1,+}}{{\bf \check c}_{i,+}}
\,\simeq\,
\frac {\Gamma[-\hat \nu_{i}]\Gamma[-\check \nu_{i+1}]} {2 \pi} 
|B \phi_{\text{BC}}|^{2(\check \nu_i  + \hat \nu_i + 1)}
e^{-\frac{8 \pi^2}{3}\frac{\Delta V_B}{H^4}}, 
\ee
where we assumed $\check \nu_{i} \simeq \check \nu_{i+1}$ everywhere but the two $\Gamma$-functions, took all $|B \phi|_{\text{BC}}$ equal in the sub-exponential factors, equated all $H(0)$ and neglected $\epsilon^2$ in the exponent. Note that the derived expressions are not valid at large $|\nu|$, 
where the asymptotic expansion~(\ref{eq:Dasymp}) is not reliable. 

\begin{figure}
\centering
\includegraphics[width=.42\textwidth]{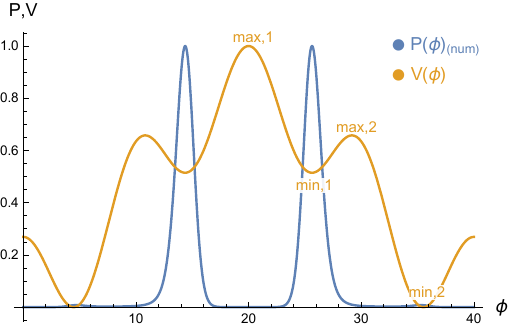}
\hspace{0.2cm}
\includegraphics[width=.42\textwidth]{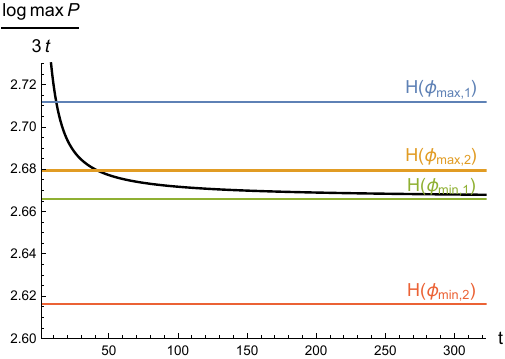}\\
\vspace{0.5cm}
\hspace{0.cm}
\includegraphics[width=.43\textwidth]{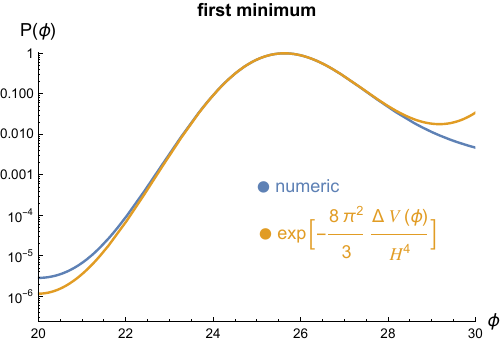}
\hspace{0.2cm}
\includegraphics[width=.45\textwidth]{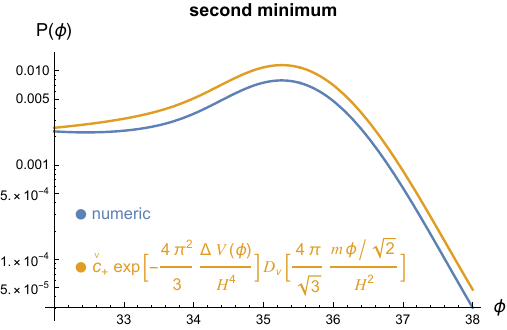}
\caption{Comparison of analytic estimates with the numeric results for the potential of Eq.~(\ref{eq:genscanV}) with $m_P=30, F=40, f=10, M^4=20, \mu^4=10, \Lambda_0 = 0.76 \times 10^3$ (in arbitrary units), $\beta = 1/2$ and a $\pi$ phase shift in one of the $\cos$ arguments.
{\bf Upper left:} scalar potential (orange, rescaled and shifted to fit the range (0,1)) and normalized field distribution $P$ at large time (blue) close to the stationary regime, obtained from a uniform initial condition.
{\bf Upper right:} numerically obtained expansion rate at the peak of the distribution as a function of time $t$ (black), and the local Hubble rates in minima and maxima. The observed expansion rate approaches the local Hubble rate in the highest minimum $\phi_{\text{min},1}$.
{\bf Lower left:} numerically obtained normalized probability distribution around the first (highest) minimum $\phi_{\text{min},1}$ (blue), compared to the asymptotic expression given by Eq.~(\ref{eq:gensolasym}) for $\nu = 0$ (orange). $\Delta V(\phi) \equiv V(\phi)-V(\phi_{\text{min,1}})$.
{\bf Lower right:} numerical result for the field distribution around the second (lowest) minimum (blue), compared to the analytic prediction of Eq.~(\ref{eq:gensol}) (orange) for $\nu = -(9/2) (H^2(\phi_{\text{min}, 1})-H^2(\phi_{\text{min}, 2}))/m^2$, $\Delta V(\phi) \equiv V(\phi)-V(\phi_{\text{min,2}})$, $\check c_+ = {\check P_2(0)}/{\check P_1(0)}$ (with the latter fixed by Eq.~(\ref{eq:Pminmin})) and $\check c_-=0$.
}
\label{fig:numstoch}
\end{figure}

Let us now discuss the matching around the few highest quadratic patches.
The $\phi\to -\phi$ symmetry implies ${\bf \hat c}_- = {\bf \hat c}_+$ for the global maximum. 
In order to satisfy the boundary condition between the global maximum and the first minimum we will choose $\check \nu \simeq 0$ in the minimum. 
In this case the left BC of the minimum reads
\bea\label{eq:match1}
\left(\frac {8\pi^2} {H^3} \right) \check J/ \check P
&\underset{\begin{subarray}{c}
\check \nu \to 0 \\
\phi\to -\infty
\end{subarray}}{\longrightarrow}&
\frac {1}{|\phi|} \left\{
 \frac 9 4 \frac{\hat H(0)^2 - \check H(0)^2}{|m^2|}  
\right\},
\eea
so that equating it with the $J/P$ of the maximum~(\ref{eq:jpmax}) we simply obtain Eq.~(\ref{eq:dtmatch}), hence it is automatically satisfied. 
According to Eq.~(\ref{eq:defnu}) the condition $\check \nu \simeq 0$ fixes $H_s$ to be close to the local Hubble parameter in the first minimum
\be
H_s^2 \simeq \check H(0)^2.
\ee
Note that applying Eq.~(\ref{eq:Pminmin}) to estimate the relative probability drop between the first minima with a small $\check \nu$ we find the factor 
\be\label{eq:gammaexp}
\Gamma[-\check \nu_{i+1}] \simeq - \frac{1}{\check \nu_{i+1}} \simeq \frac 1 9 \frac{m^2}{H_s (H_s - \check H_{i+1}(0))},
\ee
giving the same type of enhancement as was found in the discrete case~(\ref{eq:Precur}).

The BC between the first minimum and the second maximum is then satisfied as long as ${\bf \check c}_-/{\bf \check c}_+$ is suppressed as was previously estimated.
For the subsequent quadratic patches, the pattern with suppressed ${\bf c}_-$ discussed in the first part of this section applies.

Note that these results only hold for small $(\hat H(0)^2-\check H(0)^2)/|m^2|$. Let us briefly describe what happens in the opposite limit. In that case, $\check \nu = 0$ in the first minimum would imply $\hat \nu > 0$ in the global maximum, which in turn leads to a negative probability $\hat P$ at $|B\phi | \gtrsim 1/\hat \nu$. On the other hand, it can be verified that one can choose $\hat \nu \simeq 0$ to satisfy the boundary conditions between the global maximum and the first minimum. Furthermore, in this case $\hat \nu \simeq 0$ in the global maximum implies a higher $H_s$ than $\check \nu \simeq 0$ in the first minimum. Hence $\hat \nu = 0$ in the global maximum defines the highest expansion rate for the stationary regime.

Comparison of the analytic results obtained here with a numerical integration is shown in Fig.~\ref{fig:numstoch}, for a simple periodic potential with two levels of minima, satisfying $\epsilon \ll 1$ and $(9/2)(H(0)^2_{\text{max,1}}-H(0)^2_{\text{min,1}})/m^2 < 1$. The presented plots show that the expansion rate at large times approaches the local Hubble rate in the highest minimum, as predicted above. They also demonstrate a good agreement with the predicted shapes of the distributions and the relative probability drop between the two levels of minima~(\ref{eq:Pminmin}).

\section{Reheating From Bubble Collisions}\label{sec:bubreh}

Investigating the typicality of the observed Higgs mass we should also make sure that other features of the universe that we observe are typical in the analysed scenarios. In particular, we would like the ordinary observers to find themselves in  large and homogeneously reheated universes as we do. This would be the case if their space patches pass through the stage of the slow-roll inflation just before entering their vacuum. However, in principle it can also happen that the final event leading to the anthropic vacuum is a bubble nucleation. This can be the case, since after the end of the slow-roll inflation, when the overall vacuum energy of the $\phi-\chi$ landscape drops, some of the vacua still have a CC above the anthropic value, up to $\sim M^4_{\phi,\chi}$. These vacua can decay down to the anthropic range, reheating the pocket universes through the bubble wall collisions. Resulting universes would look very different from ours~\cite{Guth:1980zm,Hawking:1982ga,Guth:1982pn} and hence this process must be suppressed. Let us discuss its relevance for each of the two measures we analysed.

Let us start with the volume-weighted case.
The rate of creation of the $i$-th vacuum pocket universes from the late bubble nucleations is given by
\be\label{eq:dotPbubble}
\dot P_{i, \text{bubble nuc.}} = \left[ \Gamma_{\downarrow} P_{i-1} \right]_ {\text{after infl.}},
\ee
where we only considered one scanning field case for simplicity.
The rate of creating the universes via the normal reheating is simply given by the time derivative of Eq.~(\ref{eq:Precur})
\be
\dot P_{i, \text{slow roll}} 
= \partial_t \left[\frac{\Gamma_{\downarrow}}{3 (H_s - H_i)} P_{i-1} \right]_{\text{during infl.}}
= \left[\frac{H_s}{H_s - H_i} \Gamma_{\downarrow} P_{i-1}  \right]_{\text{during infl.}},
\ee
since all $P_{i}$ accumulated during inflation is continuously transferred into the reheated vacua, within the patches of the universe where the slow-roll inflation happens~\footnote{The rate $\dot P_{i, \text{slow roll}} $ has to contain the factor reflecting the fraction of the overall space where the inflation sector starts descending down to its minimum. However this factor enters both $\dot P_{i, \text{slow roll}}$ and $\dot P_{i, \text{bubble nuc.}}$ hence we omit it.}.
Assuming that $\left[P_{i-1} \right]_ {\text{after infl.}}$ in Eq.~(\ref{eq:dotPbubble}) equals $\left[P_{i-1} \right]_ {\text{during infl.}}$, i.e. it is refilled by the same continuous transfer mentioned above, one finds that $\dot P_{i, \text{bubble nuc.}}$ rate has a relative factor of $\Gamma_{\downarrow (\text{after infl.})}/\Gamma_{\downarrow(\text{during infl.})}$ compared to $\dot P_{i, \text{slow roll}}$. This factor is exponentially suppressed due to the Hubble scale drop after inflation, see Eq.~(\ref{eq:HMbounce}). 
Additionally, the pocket universes inside the nucleated bubbles have negative curvature, which is expected to suppress the structure formation similarly to large cosmological constant~\cite{Freivogel:2005vv,Vilenkin:1996ar,Garriga:1998px}.
Furthermore, the efficient reheating via the bubble walls collision requires two bubbles to be produced almost simultaneously within the same Hubble volume, which implies an additional $\Gamma_{\downarrow (\text{after infl.})}$ factor suppression, where
\be
\Gamma_{\downarrow (\text{after infl.})}\sim \exp\left[-\frac{8 \pi^2}{3} \frac{\Delta V_{\text{last}}}{H^4_{\text{last}}}\right],
\ee 
with $H_{\text{last}}^2 m_P^2 \simeq \frac{8 \pi}{3} M_{\chi,\phi}^4 (f_{\chi,\phi}/F_{\chi,\phi})$ and $\Delta V_{\text{last}} \simeq \mu_\chi^4, \mu_\phi^2 v_{SM}^2$ for $\chi$ and $\phi$ respectively.
If instead the two bubbles are not produced at about the same time, only a small fraction of the inner universe in the bigger bubble is expected to be reheated.
Summing up these factors we conclude that the observers created via bubble collisions are unlikely in our model.

Let us now turn to the local measure, and consider a similar sequence of events.
After the slow-roll inflation ends and the vacuum energy drops, an order one fraction of the $\phi$-$\chi$ landscape remains far above the anthropic value of the cosmological constant. Although these vacua are not habitable, all of them will eventually decay to lower-CC states. Note that since the tunnelling rate in the landscape is biased towards $\phi$ tunnelings over $\chi$, the dominant fraction of probability distribution will cross the $\phi_*$ point before reaching the anthropic line, and roll into AdS minima, see Fig.~\ref{fig:4}.   The remaining fraction will however manage to get into a vacuum with an anthropic value of CC.  Although any worldline that reached the last layer of vacua just above the anthropic line will eventually enter an anthropic-CC bubble with a probability $\sim 1$, the probability to encounter the two-bubble collision, needed for reheating, is suppressed by $\sim  \Gamma_{\text{last}}$.  
Again, as before, the number of observers can possibly be additionally suppressed due to the curvature effects~\cite{Freivogel:2005vv,Vilenkin:1996ar,Garriga:1998px}. 
We have checked numerically that even demanding the probability of normally reheated vacua to be greater than $\Gamma_{\text{last}}$ leaves the allowed parameter space presented in Section~\ref{sec:locmhcc} unaffected.


\bibliographystyle{JHEP} 
\bibliography{biblio} 

\end{document}